\documentclass{article}

\usepackage{PRIMEarxiv}

\usepackage[utf8]{inputenc} % allow utf-8 input
\usepackage[T1]{fontenc}    % use 8-bit T1 fonts
\usepackage{url}            % simple URL typesetting
\usepackage{booktabs}       % professional-quality tables
\usepackage{amsfonts}       % blackboard math symbols
\usepackage{nicefrac}       % compact symbols for 1/2, etc.
\usepackage{microtype}      % microtypography
\usepackage{lipsum}
\usepackage{fancyhdr}       % header
\usepackage{graphicx}       % graphics
\usepackage{rotating}
\usepackage{gensymb}
\usepackage{siunitx}

\graphicspath{{media/}}     % organize your images and other figures under media/ folder

\title{Self-propelling colloidal finite state machines
}
\author{
  Steven van Kesteren\textsuperscript{a}, Laura Alvarez\textsuperscript{a,b}, Silvia Arrese-Igor\textsuperscript{c},Angel Alegria\textsuperscript{c}, Lucio Isa\textsuperscript{a} \\
  a. Laboratory for Soft Materials and Interfaces,
  Department of Materials, ETH Zurich, Zurich, Switzerland \\
  b. CNRS,
  Univ. Bordeaux, CRPP, UMR5031, 33600 Pessac, France,\\
  c. Centro de Física de Materiales (SCIC-UPV/EHU) ,
  Materials Physics Center, 20018 San S\'ebastian, Spain \\
  \texttt{lucio.isa@mat.ethz.ch} \\
}

\begin{document}

\maketitle

\pagestyle{fancy}
\thispagestyle{empty}
\rhead{ \textit{ }}

\begin{abstract}
Endowing materials with physical intelligence holds the key for a progress leap in robotic systems. In spite of the growing success for macroscopic devices, transferring these concepts to the microscale presents several challenges connected to the lack of suitable fabrication and design techniques, and of internal response schemes that connect the materials' properties to the function of an autonomous unit. Here, we realize self-propelling colloidal clusters which behave as simple finite state machines, i.e. systems built to possess a finite number of internal states connected by reversible transitions and associated to distinct functions. We produce these units via capillary assembly combining hard polystyrene colloids with two different types of thermo-responsive microgels. The clusters, actuated by spatially uniform AC electric fields, adapt their shape and dielectric properties, and consequently their propulsion, via reversible temperature-induced transitions controlled by light. The different transition temperatures for the two microgels enable three distinct dynamical states corresponding to three illumination intensity levels. The sequential reconfiguration of the microgels affects the velocity and shape of the active trajectories according to a pathway defined by tailoring the clusters' geometry during assembly. The demonstration of these simple systems indicates an exciting route to build more complex units with broader reconfiguration schemes and multiple responses towards the realization of autonomous systems with physical intelligence at the colloidal scale. 

%In the future, additional components in our synthetic microswimmers could provide them with different responses and interparticle communication schemes to potentially reach “nature-like” sensing and collective motion.

%Contrary to their biological analogs or larger scale robotic systems, current artificial microswimmers lack self-regulation and rely on external control to adapt their motion. Finding strategies to endow them with in-built feedback schemes, which couple particle properties to sensing and self-motility to enable automation, remains an open challenge.
\end{abstract}

% If your first paragraph (i.e. with the \dropcap) contains a list environment (quote, quotation, theorem, definition, enumerate, itemize...), the line after the list may have some extra indentation. If this is the case, add \parshape=0 to the end of the list environment.
\section{Introduction}

As automation increasingly pervades our lives, new paradigms for autonomous systems across a broad range of length scales and application fields are in high demand. The recent upsurge of soft robotics is a clear example, where strong efforts are currently underway to endow robots with physical intelligence \cite{Sitti2021}. This concept aims at transferring functions such as perception, control and response from computational units to physical agents toward the realization of robotic materials, i.e. materials that can perform one or more of these functions without external logic and that can be integrated in autonomous devices \cite{Truby2021}. A simple guidance to constructing these systems can be based on the concept of a finite state machine (FSM). This model describes the behavior of a system as being defined by a finite number of different internal states connected by reversible transitions triggered in response to a defined set of inputs. An FSM description perfectly befits the scope of a robotic material, in the sense of a material whose function depends on its state, e.g. solid/liquid, expanded/contracted, soft/stiff, conductive/non-conductive, etc., and which can be switched by a specific trigger, e.g. temperature, light or electromagnetic signals. \\

Even if the implementation of these concepts is rapidly becoming more frequent at the macroscopic scale \cite{Breger2015,Han2017,Dou2019}, its extension to the (sub)-micrometric or colloidal scale still presents significant challenges. Current  microscale systems focus on self-propelling particles, aka artificial microswimmers or active colloids. These systems comprise units that can spontaneously convert an energy input, either externally supplied, e.g. in the form of an electromagnetic or acoustic field, or internally accessible, such as chemical fuel, into directed motion or self-propulsion. This kind of motility is at the core of the current and future applications of microswimmers to act as delivery vehicles \cite{DeAvila2017,Hortelao2021,Akolpoglu2022,Sridhar2022}%\cite{Bishop2019}
, active mixers\cite{Schuerle2019}, or remediation agents\cite{Wang2016,Wang2017} , to name a few. Crucially, self-propulsion is enabled by the fact that active colloids have an intrinsic asymmetry in their geometrical or compositional properties, which defines the directionality of their motion \cite{Ni2017,Dou2019}. However, this asymmetry is typically fixed during synthesis and fabrication\cite{Wang2019}, implying that synthetic microswimmers only have one internal state, and that propulsion is regulated by controlling the energy input.

The application of an FSM model to an active colloidal material requires the following features: i) a self-propulsion mechanism that depends on the particle's state, i.e. shape or internal properties; ii) the possibility to reversibly switch between different states in response to well-defined inputs; iii) the possibility to encode the states and their transitions in a deterministic fashion during fabrication.

In this work, we fulfill these requirements by fabricating artificial microswimmers constituted by colloidal clusters or ''molecules''\cite{Soto2014,Niu2017,Lowen2018}, which combine responsive and non-responsive particles\cite{Alvarez2021}. The clusters are deterministicly prepared by means of sequential capillarity-assisted particle assembly (or sCAPA)\cite{Ni2016}. They furthermore exhibit self-propulsion under spatially uniform AC electric fields in the kHz region, which are applied perpendicularly to the electrodes over which the particles move thanks to asymmetric electrohydrodynamic flows (EHDFs) \cite{Ristenpart2007, Ma2015, Ma2015a,Shields2017} . The transition between the different dynamical states is triggered by externally-imposed temperature signals, which cause the reconfiguration of the shape and dielectric properties of the clusters. We first begin by conceptually describing our approach to achieve microswimmers with multi-state dynamics and then focus on their experimental realization and characterization.

\subsection{Self-propelling colloidal finite state machines: the concept}\label{sec:sectA}

The conceptually simplest case of a self-propelling particle with multiple dynamical states is a two-state particle. Given the requirement for asymmetry, a two-state active particle can be realized by constructing a dumbbell where one of the two lobes changes its properties in response to an input signal. 

By selecting EHDF-induced propulsion and temperature signals, dumbbells comprising one non-responsive, e.g. polystyrene (PS), lobe and one responsive, e.g. Poly(N-isopropylacrylamide)-based, lobe fulfill the criteria. Poly(N-isopropylacrylamide) (PNIPAM) is a thermo-responsive polymer that undergoes a reversible lower critical solubility transition at a well-defined temperature. PNIPAM-based particles (or microgels) correspondingly undergo a swelling/deswelling transition at that temperature, called the volume phase transition temperature or VPTT. Upon crossing the VPTT, microgels experience a volumetric change coupled to a variation of their dielectric properties \cite{Su2014,Mohanty2016,Alvarez2021,Nieves2000}. Self-propulsion generated by EHDFs in dumbbells depends both on the relative dimensions of the two lobes and on their dielectric properties \cite{Ma2015,Ni2017}. The existence of two different dynamical states for reconfigurable, PNIPAM-based active dumbbells has in fact been recently demonstrated \cite{Alvarez2021}.   

However, the VPTT of the microgels depends on their chemistry, their internal architecture, e.g. crosslinking density \cite{Ruscito2020}, and the solvent properties \cite{Scherzinger2014}. In this work, we synthesize and use two types of thermo-responsive PNIPAM-based microgels to enable the access of multiple dynamical states (see Methods). In particular, we use two different co-monomers, methacrylic acid (MAA) and acrylic acid (AA), in the polymerization of PNIPAM to obtain microgels with different VPTTs. The PNIPAM-co-MAA microgels have a green-fluorescent (BODIPY) core and the PNIPAM-co-AA microgels have a red-fluorescent (Nile-red) core, and are henceforth termed green (G) and red (R) microgels. Due to the higher hydrophilicity of AA compared to MAA, its VPTT is shifted to higher temperatures \cite{Lin2006, Kratz2000,Gao2013}. The VPTT of green-core microgels is $T_{G \rightleftharpoons G'} \approx$ 29 \degree C and the one of the red-core microgels is $T_{R \rightleftharpoons R'} \approx$ 33 \degree C (see Supplementary Section 1 for further details).

\begin{figure}
\includegraphics[width=\linewidth]{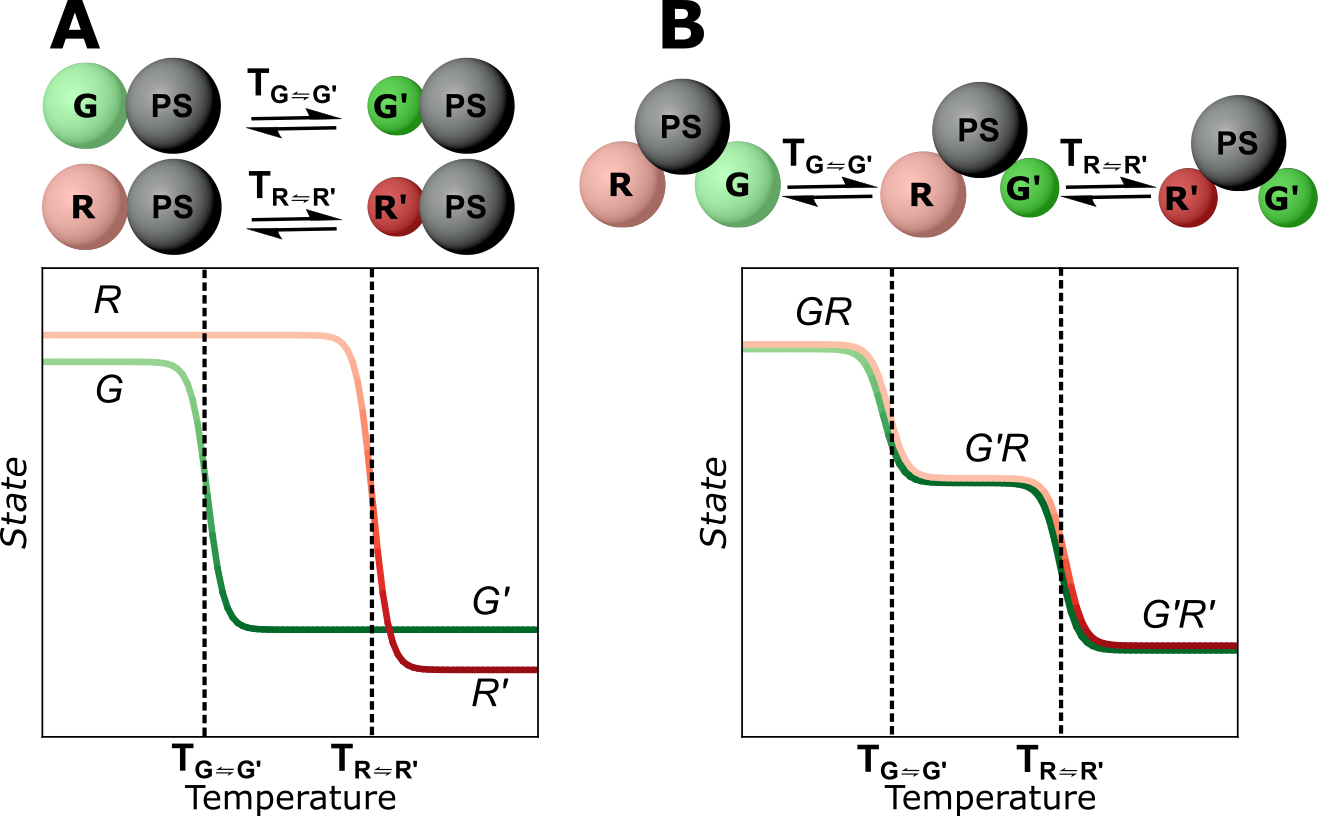}
\caption{\label{fig:Fig1}  \textbf{Schematic representation of the FSM model for reconfigurable colloidal clusters.} A, Representation of the reconfiguration of the of G-PS and R-PS dumbbells upon crossing the respective transition temperatures $T_{G \rightleftharpoons G'}$ and $T_{R \rightleftharpoons R'}$ and schematic of the corresponding transition between the different states as a function of temperature $T$. B, Extension of the previous concept to a three-state particle (R-G-PS trimer).  }
\end{figure}

By using these microgels, we can therefore construct two different types of dumbbells, G-PS and R-PS, respectively, of which we can schematically represent the different states and their transitions as shown in Figure \ref{fig:Fig1}A. The G-PS dumbbells undergo a transition between a swollen state $G$ to a collapsed state $G'$ ($G \rightleftharpoons G'$) at a temperature $T_{G \rightleftharpoons G'}$, and correspondingly the transition $R \rightleftharpoons R'$ takes place at $T_{R \rightleftharpoons R'}$. As it will be showed in Section \ref{sec:sectB}, these two transitions correspond to a switch between two different self-propulsion modes, with a an adaptation of the propulsion velocity $V_G \rightleftharpoons V_{G'}$ and $V_R \rightleftharpoons V_{R'}$, respectively.

The existence of two different transition temperatures opens the way to the fabrication of microswimmers with three distinct dynamical states by incorporating both G and R microgels in a colloidal trimer together with a PS particle, as schematically shown in Figure \ref{fig:Fig1}B. Upon crossing $T_{G \rightleftharpoons G'}$, the trimer goes from a fully swollen state to one where only the G microgel is collapsed $GR \rightleftharpoons G'R$; when $T_{R \rightleftharpoons R'}$ is crossed, both microgels collapse and the trimer undergoes the $G'R \rightleftharpoons G'R'$ transition. The order of the transitions is reversed by progressively reducing temperature. In the case of the trimer, because the collapse of one microgel can cause the breaking of the axial symmetry of the cluster, in addition to three distinct translational velocities $V_{GR} \rightleftharpoons V_{G'R} \rightleftharpoons V_{G'R'}$, three different angular velocities can also be accessed $\omega_{GR} \rightleftharpoons \omega_{G'R} \rightleftharpoons \omega_{G'R'}$.

\begin{figure*}
\includegraphics[width=\linewidth]{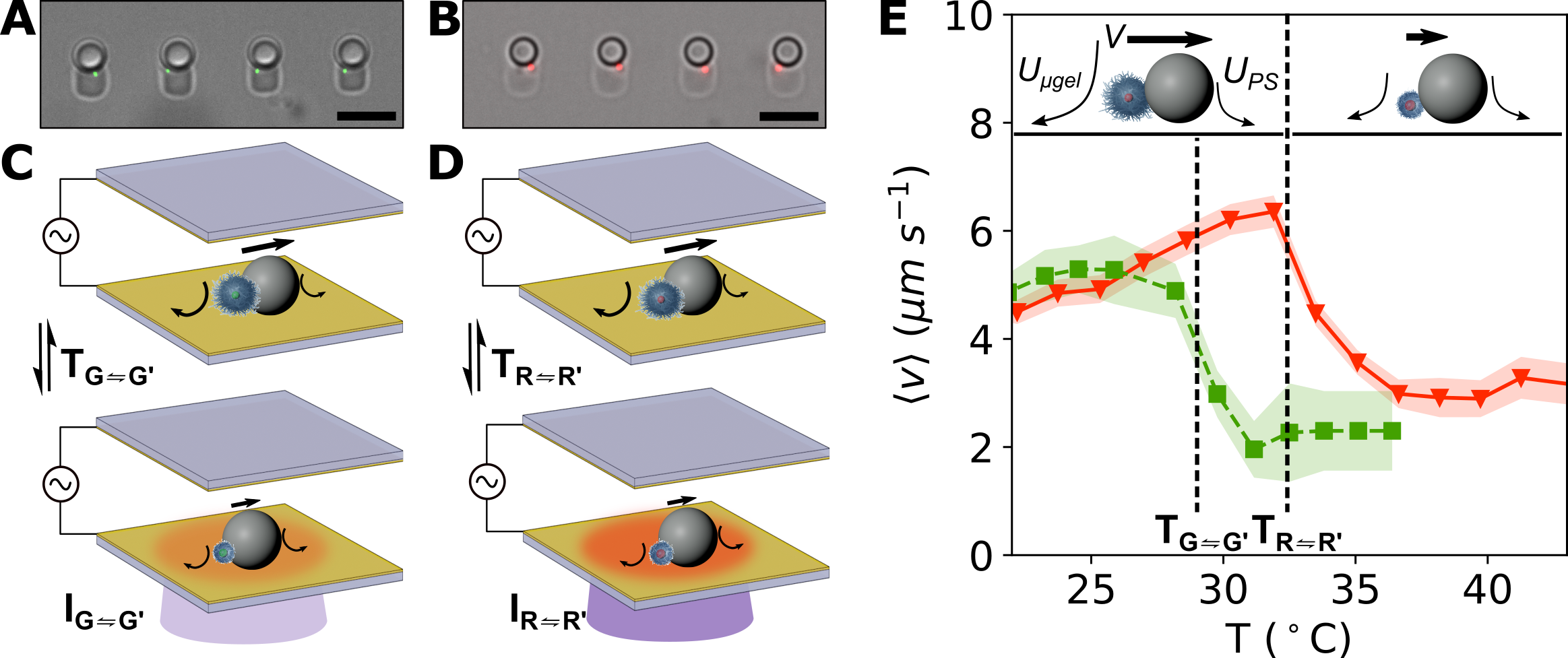}
\caption{\label{fig:Fig2} \textbf{Two-state microswimmers.} A-B, Combined bright-field+fluorescence micrographs of G-PS (A) and R-PS (B) dumbbells in the sCAPA traps.  The scale bars are 5 \textmu m. C-D, Schematic of the experiment for the two-state temperature-induced switching in the self-propulsion of G-PS (C) and R-PS (D) dumbbells. The curved arrows indicate the EHDFs generated by the PS and microgel lobes, respectively, and the straight arrows represent the net propulsion due to these EHDFs. Plasmonic heating of the substrate illuminated with a light of intensity $I_{G \rightleftharpoons G'}$ ($I_{R \rightleftharpoons R'}$) raises the local temperature above $T_{G \rightleftharpoons G'}$ ($T_{R \rightleftharpoons R'}$) inducing a phase transition in the microgels. Note that because $T_{R \rightleftharpoons R'} > T_{G \rightleftharpoons G'}$, the corresponding illumination intensity is higher. E, Mean instantaneous self-propulsion velocity $<v>$ as a function of temperature $T$ for the R-PS (triangles) and G-PS (squares) dumbbells. The shaded bands are the 99\% confidence intervals calculated over 350 and 70 particles for the R-PS and G-PS, respectively . The vertical dashed lines mark the transition temperatures for the green and red core microgels, respectively. The inset schematically shows the switching of the self-propulsion state upon crossing the critical temperature, taking a R-PS dumbbell as an example. The curved arrows represent the EHDFs generated by the microgel and the PS particle, with characteristic velocities \textit{U\textsubscript{\textmu gel}} and \textit{U\textsubscript{PS}}, respectively, and the straight arrows indicate the net propulsion with velocity \textit{V}.}
\end{figure*}

\subsection{Two-state dynamics: self-propelling responsive dumbbells}\label{sec:sectB}  

After introducing the concept in the previous Section, we now show that the behavior schematically displayed in Figure \ref{fig:Fig1}A can be experimentally implemented. We synthesized PNIPAM-co-MAA (G) microgels and PNIPAM-co-AA (R) microgels that have a hydrodynamic diameter of $\approx$ 2-3 \textmu m before and $\approx$ 0.7-1.2 \textmu m after their respective VPPT (Figure S4). In both cases the fluorescent cores are made from poly 2-trifluoromethylacrylic (PTFMA) and have a diameter of around 200 nm, to visualize the different microgels under the microscope. The G-PS and R-PS dumbbells are obtained using sCAPA, as previously described \cite{Ni2016,Alvarez2021}. In short, both PS particles and microgels are sequentially deposited into pre-designed traps microfabricated on a polydimethylsiloxane (PDMS) template. An evaporating droplet of the target particle suspension is moved over the PDMS template at a controlled speed and capillary forces selectively place the particles inside the traps. PS particles are deposited first, followed by either the G or R microgels to obtain the dumbbells inside the traps (see Figure \ref{fig:Fig2}A-B). After deposition, the particles are linked by thermal sintering at 75\degree C for 15 minutes and harvested using an adhesive sacrificial layer of glucose.

The resulting dumbbells are dispersed in a dilute HEPES buffer and confined between two transparent electrodes consisting of glass slides coated by a Cr and a Au layer covered with a polyethylene glycol (PEG) brush (see Methods). The electrodes are separated by a 240 \textmu m spacer and connected to a function generator that applies an AC electric field of 7 V peak-peak amplitude (~300 V cm\textsuperscript{-1}) and 800 Hz frequency perpendicular to the electrodes. Upon application of the AC field, both the electrodes and the particles polarize; the presence of the particles close to the bottom electrode (where they sediment) causes a local distortion of the electric field generating a component that is tangential to the substrate, which in turns leads to the recirculation of ions, the so-called electrohydrodynamic flows (EHDFs), around each colloid \cite{Ristenpart2007}. At a fixed voltage and frequency, the magnitude of the EHDFs depends on the particle size and dielectric properties \cite{Ni2017, Ma2015,Ristenpart2007}. However, for homogeneous spherical particles, the EHDFs are radially symmetric and do not induce any net propulsion. Conversely, for dumbbells with lobes of different size and dielectric properties, the symmetry of the EHDFs is broken and a net fluid flow is generated, which leads to self-propulsion. In particular, if $U_{i}$ is the magnitude of the velocity of the EHDF generated around particle $i$, the propulsion velocity of the dumbbell can be written, in first approximation, as $v_{ij}=U_{i}r_{j}+U_{j}r_{i}/(r_{i}+r_{j})$, where $r_{i}$ is the radius of particle $i$ (or $j$, respectively) \cite{Ma2015}. Given that both $U_{PS}$ and $r_{PS}$ do not strongly depend on temperature, a sudden change in propulsion velocity is expected upon crossing $T_{G \rightleftharpoons G'}$ and $T_{R \rightleftharpoons R'}$ for the G-PS and R-PS dumbbells, respectively, because both the size and the dielectric properties of the microgels switch between different states upon crossing their respective VPTT (see Supplementary Section 2 for further details). 

Experimentally, we induce controlled and rapid temperature variations by exploiting the plasmonic heating of the Au layer within the electrode upon illumination at 395 nm and 555 nm \cite{Son2015}. By adjusting the light intensity, we can locally heat the system above $T_{G \rightleftharpoons G'}$ or $T_{R \rightleftharpoons R'}$ and correspondingly reconfigure the dumbbells (\ref{fig:Fig2}A-B; (see Supplementary Section 3 for further details) for an experimental calibration and a numerical verification of the heating via finite element modelling (FEM) with COMSOL Multiphysics \cite{COMSOL}). The system equilibrates within seconds and cools down upon removal of the illumination to ensure the reversible transition between the different states.   

In line with previous observations \cite{Alvarez2021}, the collapse of the microgels significantly reduces the magnitude of the propulsion velocity of the microgel-polystyrene dumbbells (Figure \ref{fig:Fig2}C). In particular, $T_{G \rightleftharpoons G'}$ or $T_{R \rightleftharpoons R'}$  are sufficiently separated so that G microgels collapse while R ones are still in the swollen state, resulting in reconfiguration and velocity changes at distinct temperatures. This response follows the scheme proposed in \ref{fig:Fig1}A and holds the key to access additional dynamical states when combining both microgels in one active cluster. \\

\begin{figure}
\centering
\includegraphics[width=0.5\linewidth]{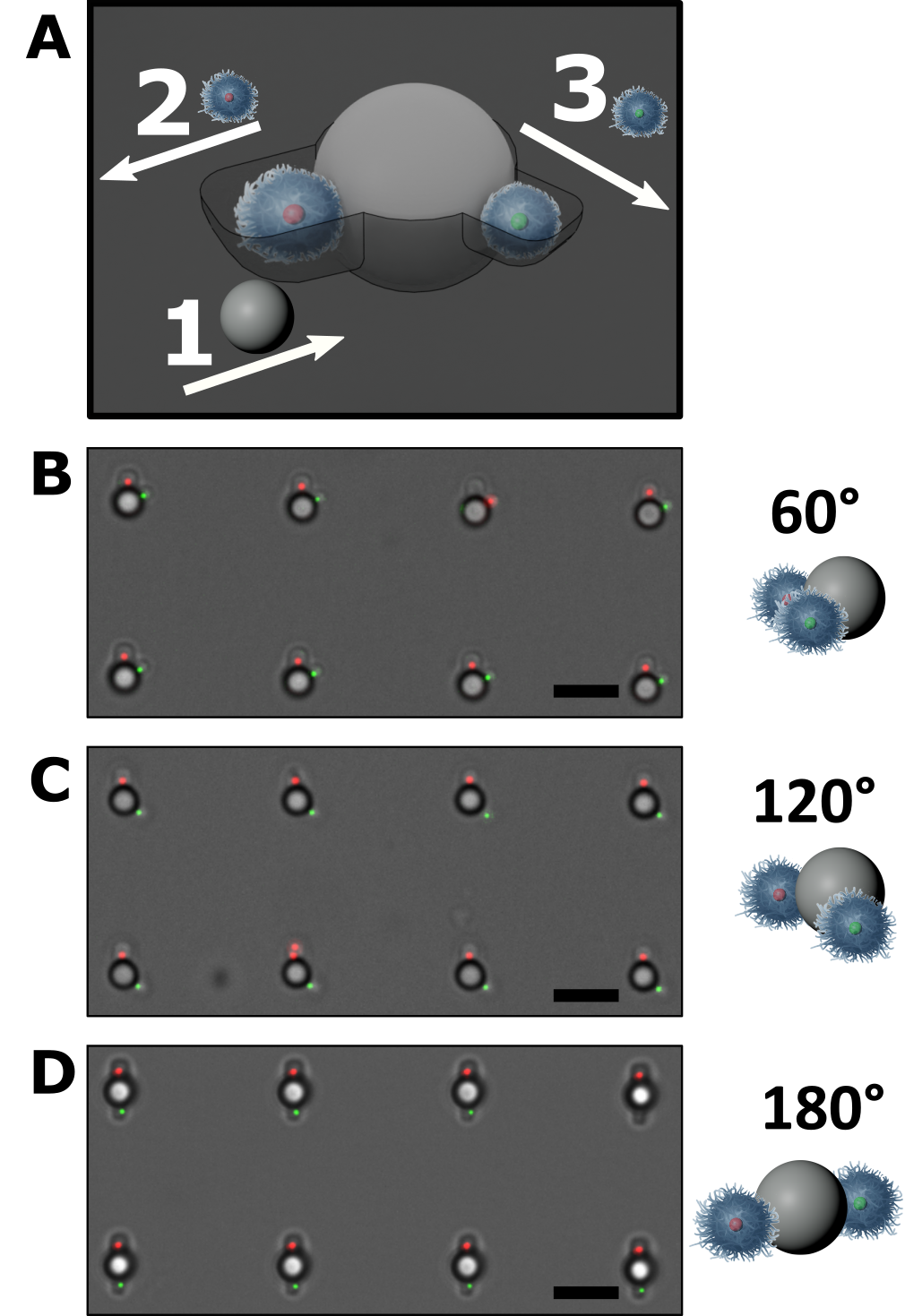}
\caption{\label{fig:Fig3} \textbf{Assembly of reconfigurable trimers.} A,  Schematic showing the filling of the 3D-traps with sCAPA for a single colloidal 120-cluster. The numbers indicate the order of the deposition steps and the arrows respective deposition directions. B-D, Combined bright-field+fluorescence micrographs of R-G-PS trimers in the sCAPA traps with 60 (B), 120 (C) and 180 (D) degrees between the microgels. Scale bars are 5 \textmu m}
\end{figure}

\subsection{Assembly of multi-microgel colloidal clusters}
However, in order to realize the system proposed in Figure \ref{fig:Fig1}B, we need to achieve deterministic control on the relative position of the two microgels in the active G-R-PS trimer. As it can be seen in \ref{fig:Fig2}A-B, conventional sCAPA in rectangular traps does not have sufficient positional control for the microgels, which are deposited on either side of the PS particles. In order to overcome this limitation, we have recently demonstrated a further refinement of sCAPA that exploits traps with three-dimensional height profiles realized by direct writing using a two-photon nanolithography system (Nanoscribe GT2 Professional), as schematically shown in Figure \ref{fig:Fig3}A-B (see Methods and Supplementary Section 4 for further details).\cite{vanKesteren2022} We use the different height of the different sections of the trap to guide the deposition of the polystyrene particles and the R and G microgels into trimers with well-defined opening angles. The assembly process requires careful tuning of the trap geometry as well as of the deposition medium, speed, direction and order \cite{Ni2015}. In particular, the PS particles are deposited first, followed by the R and then the G microgels (see Figure \ref{fig:Fig3}B). By changing the geometry of the traps, we produced three-particle clusters with opening angles of 60, 120 and 180$^{\circ}$ degrees. We refer to these systems as 60-, 120-, or 180-clusters and examples are shown in Figure \ref{fig:Fig3}C. The deposition yields go up to 75\% for the 120-clusters (see Supplementary Section 5 for further details) and we use the same linking and harvesting procedures as for the dumbbells.

\begin{figure*}[h]
\includegraphics[width=\linewidth]{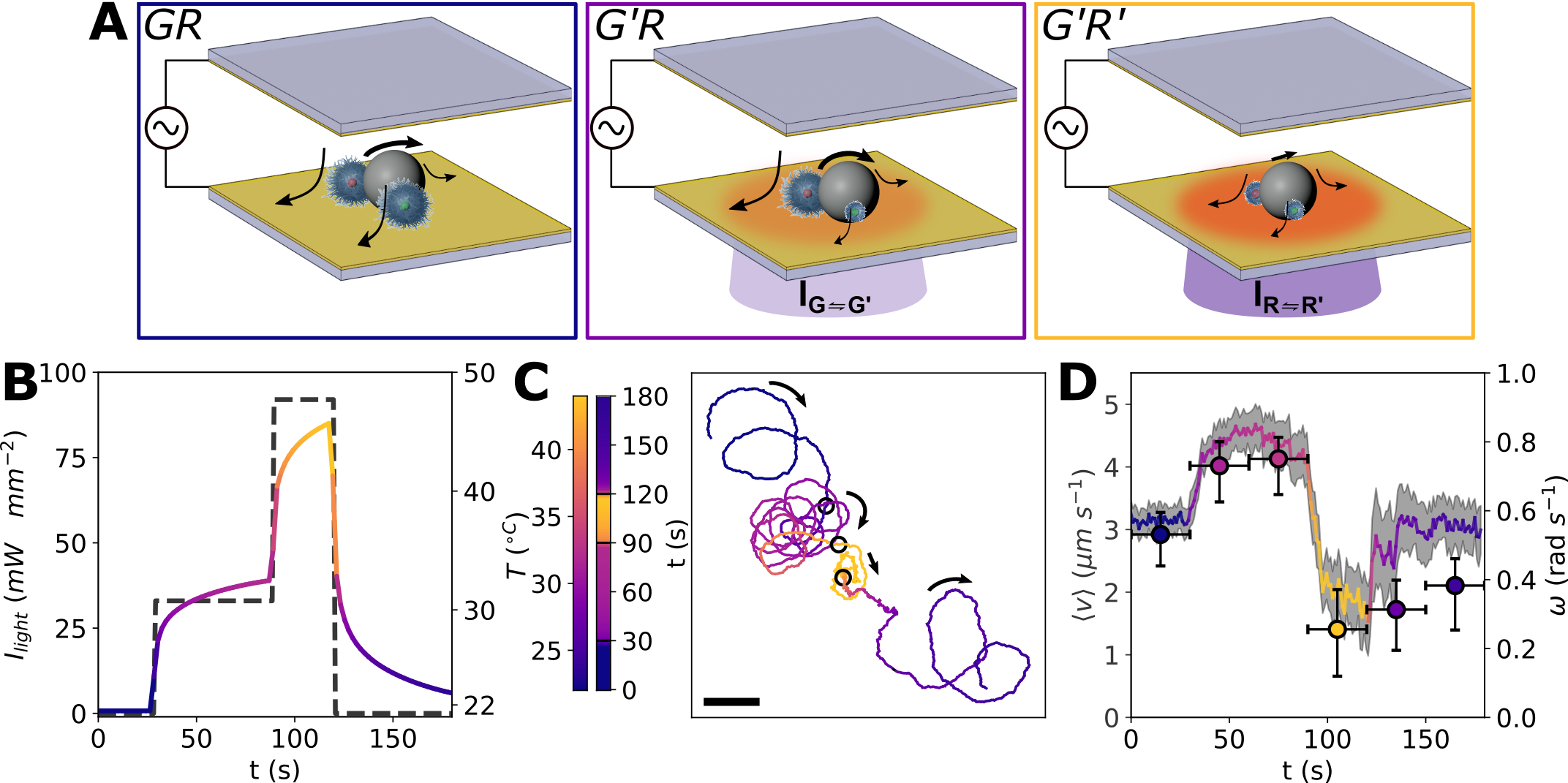}
\caption{\label{fig:Fig4} \textbf{Three-state dynamics for G-R-PS 120-clusters.} A, Schematic illustrating the 120-cluster in state \textit{(GR), (G'R)} ,or \textit{(G'R')} depending on the illumination conditions. The arrows indicate the EHDFs and the net propulsion like in the previous figures. B, Light exposure sequence (dashed line) over time and resulting temperatures (solid line) as predicted by finite elements simulations. The color coding of the temperature corresponds to the color scale bar on the right of the graph.  C, Example trajectory of a G-R-PS 120-cluster color-coded with time and corresponding temperature as represented by the color scale bars on the left of the graph. The scale bar is 20 \textmu m.  D, Mean instantaneous velocity (solid line) and angular velocity (symbols) of the microswimmer over time color-coded with temperature. The shaded bands and vertical error bars are the 99\% confidence intervals calculated over 39 trajectories for the instantaneous velocity and angular velocity, respectively. }
\end{figure*}

\subsection{Three-state motility of 120-clusters}

We first focus on the dynamical response of the 120-clusters. After assembly and harvesting, the trimers are inserted in the same experimental cell that we previously described for the dumbbells and the transverse AC electric field (7 V peak-to-peak at 800Hz) is applied together with illumination of varying power density $I$ at 395 nm to induce the local heating (Figure \ref{fig:Fig4}A). In particular, we use a sequence of illumination/heating steps in the following order: 30 s no illumination ($\approx$ 21 \degree C), 60 s at 33 mW mm\textsuperscript{-2} ($\approx$ 31\degree C), 30 s at 92 mW mm\textsuperscript{-2} ($\approx$ 45 \degree C) and finally 60 s no illumination (Figure \ref{fig:Fig4}B). This sequence enables us to switch between a state where both microgels are swollen \textit{(GR)}, one where one microgel is swollen and the other one is collapsed \textit{(G'R)}, one where both microgels are collapsed \textit{(G'R')} and back to the initial state where both microgels are swollen again \textit{(GR)} (Figure \ref{fig:Fig4}B). We observe rapid, reversible switching amongst the various states, which is reflected in a qualitative and quantitative variation of the trajectories (Figure \ref{fig:Fig4}C): in state \textit{(GR)} the trajectory is curvilinear with a constant radius of $\approx 12 \mu$ m, in \textit{(G'R)} the particle continues with a strong chiral motion but the radius of curvature is significantly decreased to 5.5 \textmu m, and, finally, in \textit{(G'R'}) the propulsion velocity drops and the rotational diffusion increases, resulting in less-directed motion. Upon removal of fluorescence illumination, the trajectory resumes the original shape of state \textit{(GR)}. 

\begin{figure*}[h!]
\includegraphics[width=\linewidth]{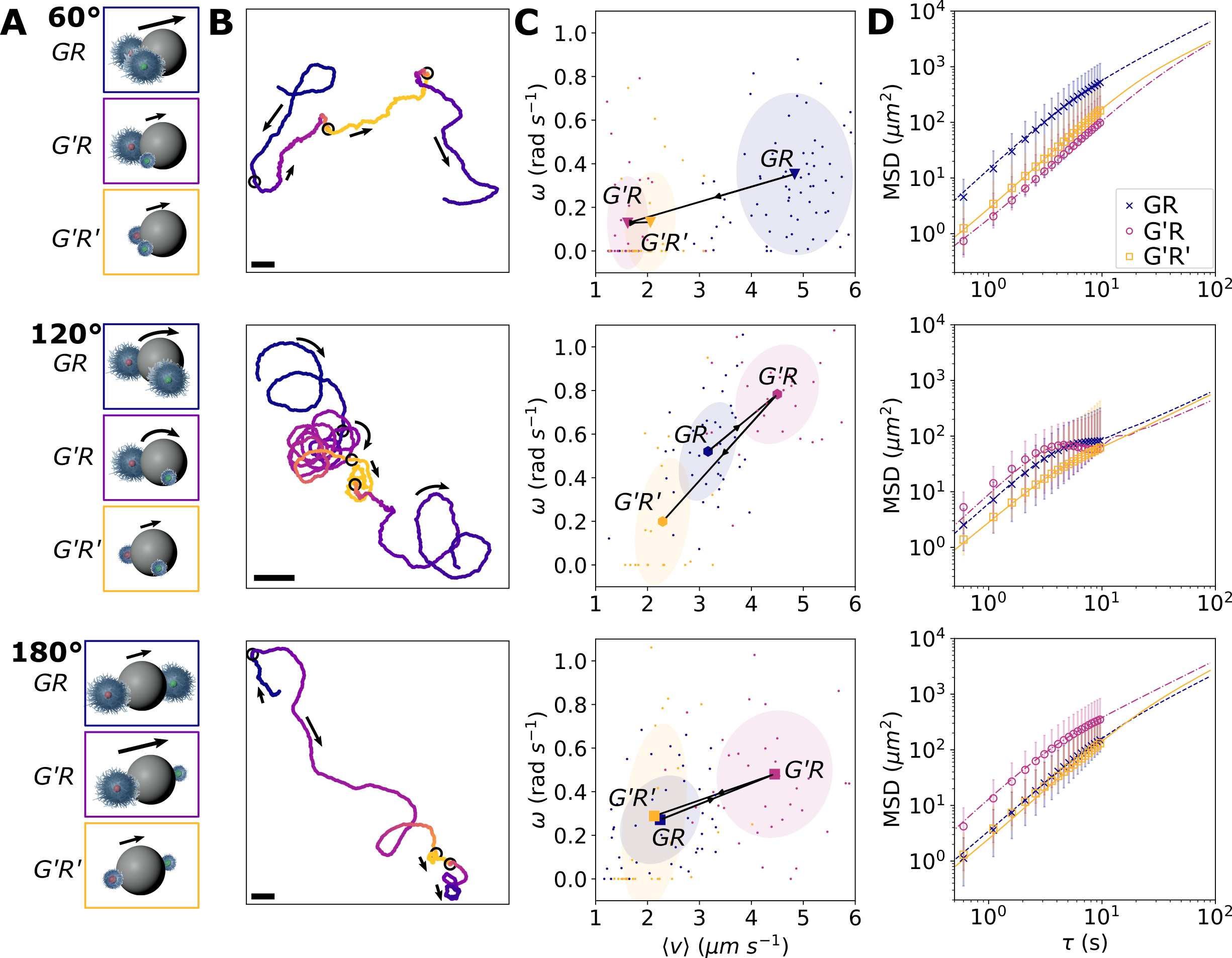}
\caption{\label{fig:Fig5} \textbf{Multi-state motility of  different active clusters. }  Each row corresponds to the 60-, 120- or 180-clusters, respectively. A, Schematic of the representative motion of the three clusters in state \textit{(GR)},\textit{(G'R)} and \textit{(G'R')}. B, Example trajectories color-coded with temperature (same temperature range as in Figure \ref{fig:Fig3}). Scale bars are 20 \textmu m . C, Scatter plot of the mean instantaneous velocity and angular velocity for each particle in \textit{(GR)},\textit{(G'R)} and \textit{(G'R')}. The elliptical shaded regions correspond to one standard deviation of both quantities calculated over 54, 39 and 45 trajectories for the 60-, 120- or 180-clusters, respectively. D, Ensemble-averaged MSD in \textit{(GR)},\textit{(G'R)} and \textit{(G'R')}. Symbols: experimental data. Dashed lines: MSD fits for a rotating active Brownian particle \cite{Archer2015}. } 
\end{figure*}

By quantitatively tracking the motion of the trimers as a function of time/illumination, we further characterize their motility in terms of their instantaneous and angular velocity. The two quantities are coupled due to the nature of the trajectory (Figure \ref{fig:Fig4}D), but they can be independently measured from the experiments (see Supplementary Section 6 for the calculation of the angular velocity) and both contribute to how the active particles explore space (see Figure \ref{fig:Fig5}). \\
The shape of the trajectories and net swimming velocity are rationalized by considering the direction of the propulsive force and of viscous drag. As we have previously discussed in Figure \ref{fig:Fig1}, in the case of dumbbells, propulsion is aligned with the direction of compositional asymmetry and, as a consequence of geometry, the center of propulsion and of hydrodynamic drag coincide. However, this in not necessarily the case for the trimers. As schematically shown in Figure S12 (Supplementary Section 7), we can rationalize the propulsion of the trimer as the linear superposition of the propulsion generated by two dumbbells, respectively connecting the PS particles to the G and the R microgels. Variations in the magnitude of each velocity vector and of the size of the respective dumbbell lobes, may cause a misalignment between the center of propulsion and the center of hydrodynamic drag, giving rise to a rotation-translation coupling and the corresponding emergence of a active torque\cite{Brenner1967,Chakrabarty2014,Zhang2019,Kraft2013}. With this simple image in mind, and considering the temperature response of each individual self-propelled dumbbell, in state \textit{(GR)}, each microgel has a similar size and generates similar EHDFs. However, these are not identical, leading to a breaking of the in-plane symmetry of the trimer's propulsion and drag, which leads to a curvilinear trajectory with a well-defined chirality and a large radius. In \textit{(G'R)}, the G-microgel collapses with a dramatic drop on its propulsion velocity,  while the propulsion velocity of the R-microgel increases slightly. The greater asymmetry between both microgels leads to an increased torque, and correspondingly to a tightening of the curvilinear radius. The \textit{G'R}-trimer's instantaneous velocity is slightly larger compared to the one in state \textit{(GR)}, due to the moderate increase of the EHDFs of the R-PS dumbbell, in conjunction with the reduced drag stemming from the collapse of the G-microgel. The combination of these factors leads to an increased angular velocity. Finally, in \textit{(G'R')}, both microgels are collapsed and provide a significantly lower propulsive force. Correspondingly, the trimer's velocity dramatically drops. This fact, combined with the overall decreased size of the microswimmer implies that rotational diffusion plays a greater role in determining the trimer's trajectory in state \textit{(G'R')}.\\

\subsection{Multi-state space exploration}

The same simplified approach can be used to describe the motility of trimers with the other internal angles (see Supplementary Section 7 for further details), where reconfiguration can be used to tune how different active clusters explore space. By examining active trajectories, we immediately observe a clear qualitative difference in the dynamical behavior amongst the 60-, 120- and 180-clusters (Figure \ref{fig:Fig5}A). 

The 60-clusters (Figure \ref{fig:Fig5}A top) have the largest instantaneous velocity in state \textit{(GR)}. However, unlike the 120-clusters, their velocity immediately drops in \textit{(G'R)} and remains similar in \textit{(G'R')}. The 180-clusters (Figure \ref{fig:Fig5}A bottom) conversely have a low instantaneous velocity in state \textit{(GR)} compared to the 120-cluster. However, the transition to state \textit{(G'R)} shows a much more significant speeding up without a large change in the angular velocity, while in state \textit{(G'R')} the 180-cluster behave very similar to the 60- and 120-clusters.  We therefore show that simple reconfiguration strongly affects the motility of the clusters in the different states.%\cite{Archer2015,Brenner1967,Chakrabarty2014,Kraft2013,Zhang2019} \\

Similarly to our description of the 120-cluster, the behaviour of the 180-cluster can be rationalized by the force vector model shown in the Supplementary Section 7. In state \textit{(GR)} the propulsive forces of the 2 microgel-PS dumbbells are similar in magnitude but oppose each other leading to a very small net propulsion. Switching to \textit{(G'R)} breaks this balance, the G microgel collapses and the clusters speed up. However, the behaviour of 60-clusters appear different, and self-propulsion essentially ceases already at $T_{G \rightleftharpoons G'}$ and does not change upon crossing $T_{R \rightleftharpoons R'}$. We hypothesize that, given their dimensions relative to the PS particle, the two microgels substantially overlap in the swollen state with an internal angle of 60$^\circ$, such that the collapse of the G microgel already causes the R microgel to collapse, practically eliminating any difference between \textit{(G'R)} and \textit{(G'R')} (see Supplementary Section 8 for further details). \\. 

Finally, the three different states of the clusters are associated with different ways to explore space, which can be represented by plotting their mean-square displacements (MSDs) \cite{Archer2015}. The 60- and 180-cluster have a clear ‘fast’ state, which is \textit{(GR)} for the 60-clusters and \textit{(G'R)} for 180-clusters. These states correspond to high instantaneous velocities and small angular velocities explore, for which space is explored significantly faster by particle ensembles than the low-velocity states at any time. The MSDs for 120-cluster are more nuanced, where mean displacements are greater in state \textit{(G'R)} at short times, but motion becomes more confined at longer times due to the strong chiral motion. The distinct dynamical behavior of different clusters in different states indicates a broad range of possibilities to develop adaptive artificial microswimmers by design.

%% Although the microswimmers in this work are primarily a fundamental proof-of-concept, these AC-field driven microswimmers can work in microfluidic devices. For example, due to their different MSDs 60-clusters are expected to accumulate in regions between 29-33 \degree C, while 180-clusters avoid those regions. Therefore, these microswimmers can be used to autonomously perform certain tasks like catalysis or transport controlled by their local environment. \cite{García2013}

\section{Conclusions and outlook}

In this manuscript we have demonstrated that the combination of deterministic assembly using sCAPA and the utilization of stimuli-responsive microgels enables the realization of synthetic microswimmers that can transition among distinct internal states, which are in turn coupled to their dynamics. Importantly, and differently from typical light- or magnetically-activated particles, where the stimuli for propulsion and the ones that regulate motility coincide \cite{Han2017,Lozano2016,Lozano2019}, our system's propulsion mechanism is orthogonal to the reconfiguration one. 

The current study focuses on demonstrating the feasibility of self-propelling finite state machines at the colloidal scale and on the characterization of the single-particle dynamics for different microswmimmer designs. However, the independence of the reconfiguration pathway enables robust control of the particle motility and the possibility to easily implement spatially and temporally-modulated motility landscapes to investigate collective effects, in analogy with investigations of the collective dynamics of light-regulated bacteria \cite{Arlt2018,Frangipane2018}. In living organisms, the adaptation to external stimuli is internally regulated and characterized by finite transition times. Conversely, synthetic microswimmers subjected to sophisticated perception and adaptation schemes\cite{Lavergne2019,Khadka2018,Franzi2020} are externally controlled and the feedback between input signals and motility regulation can be considered instantaneous over the characteristic timescales of active motion. Nonetheless, the presence of finite response times, even for external control strategies, has been shown to lead to emergent dynamical and collective effects \cite{FernandezRodriguez2020,Yang2018,Mijalkov2016}. Our colloidal FSMs fall in between these categories, where propulsion and reconfiguration are externally triggered, but where the transitions between the different states are encoded during fabrication and take place with a characteristic internal time scale. 

Given our findings, many further avenues for future work open up. For instance, the level of single-particle control may already be amenable to the demonstration of lab-on-chip devices. The AC-electrodes could be easily integrated in microfluidic systems, where, guided by thermal signals, the reconfigurable microswimmers can navigate and perform different functions. For example, we envisage that the 60-clusters accumulate in regions where the temperature is between 29-33 \degree C, corresponding to state \textit{(G'R)}, where they have a small MSD. They could therefore efficiently explore space in \textit{state GR} until they reach a region with a higher temperature and linger there, i.e. to aid or perform a chemical reaction, if carrying the appropriate load. Conversely, in the same state, 180-clusters would spend little time in these hotter regions due to a corresponding increase in their MSDs. They could thus be used to remove reaction products and transfer them to colder regions, where their motility drops. Even though thermal signaling may not be practical for all applications, incorporating responses to different stimuli may offer further options.
%most applications and a similar system sensitive to chemical signals will be more suited in practice. \cite{Garcia2013} \\
%Here we realized AC electric field actuated multi-state active collodial clusters with internal adaptive mechanism. First, we investigate the relation between temperature and propulsion of dumbbell clusters made of hard spheres and responsive microgels with different VPTT's. Subsequently, we use sCAPA to combine these different microgels and hard spheres into three-particle clusters with different internal angles. Hereby, we demonstrate unprecedented control over geometry and material composition during the assembly of the colloidal clusters thanks to the two-photon lithography fabrication technique to fabricate the colloidal molds. Rich dynamical behavior is achieved by assembling two thermoresponive PNIPAM microgels colloids and a hard sphere in simple shapes. The different internal angles of the colloidal clusters lead to different active behaviors. Small changes in their pre-designed shape and, thus, in their mobility are possible due to the reconfiguration of the microgels throughout temperature variations. 

In fact, the modular assembly enabled by sCAPA makes it possible to incorporate other soft components, which can reconfigure in response to different stimuli. PNIPAM-based microgels that are responsive to light \cite{Das2007}, pH \cite{Kratz2000,Gao2013}, redox state \cite{Suzuki2008}, magnetic fields \cite{Rittikulsittichai2016} and (bio)chemical molecules \cite{Tang2014} have been synthesized. The use of multiple, orthogonal stimuli would greatly increases the number of possible states, compared to using only one stimulus, e.g. temperature, with different transitions.  

Finally, in order to successfully assemble precisely-programmed reconfigurable clusters, we had to push the boundaries of sCAPA beyond the state of the art with the introduction of 3D height profiles within the traps to direct the deposition of soft and hard particles in deterministic shapes. In the future, we also envisage that additional strategies can be used to incorporate reconfigurable parts in microscale active units. Recent advances in 3D-printing of microscale responsive hydrogels with two-photon nanolithography could allow for more flexible fabrication of soft structures \cite{Hippler2019}, toward hybrid colloidal materials\cite{Hu2021}.
All of these efforts will take a very active community of researchers and engineers one step, or more, closer to the goal of realizing autonomous microscale systems with capabilities mirroring the ones of larger scale robotic units.

\subsection*{Acknowledgements}
LA acknowledges financial support from the European Soft Matter Infrastructure (EUSMI) proposal number S180600105. This project has received funding from the European Research Council (ERC) under the the European Union’s Horizon 2020 research and innovation programme grant agreement No 101001514. We thank Alexander Kuehne and Dirk Rommels for help with particle synthesis and discussion.

\newpage

\section{Materials and Methods}
All chemical if not state otherwise were used as provide by the supplier. Triton X-45, Acrylic acid (AA),Methacrylic Acid (MAA), N, N' Methylenebis(acrylamide) (BIS), 2,2,2-Trifluoroethyl methacrylate (TFMA), Sodium dodecyl sulfate (SDS), Potassium persulfate (KPS), BODIPY, and Nile red were purchased from Sigma-Aldrich. N-Isopropylacrylamide (NIPAM) was purchased from Sigma-Aldrich and purified by recrystallization in Toluene/Hexane 50:50. 4-(2-hydroxyethyl)-1-piperazineethanesulfonic acid (HEPES) was purchased from VWR. 

\subsection*{Synthesis of core-shell pNIPAM microgels}

The microgels with fluorescent cores and thermo-responsive shells were prepared in a two-step synthesis \cite{Go2014}:  first the PTFMA-cores containing a fluorescent dyes were prepared using free radical emulsion polymerization, then a shell of PNIPAM and PAA or PMAA co-polymer was grown by free radical precipitation polymerization around this core. For the cores, TFMA (10 mL; 12 g; 70,3 mmol), NIPAM (940 mg; 8,31 mmol), SDS (30 mg; 0,104 mmol) and Nile red (5 mg; 0.016 mmol) or BODIPY 493/503 (5 mg; 0,016 mmol) were added in water (30 mL) and stirred at 600 rpm for 20 min and purged with N\textsubscript{2}. The reaction mixture was heated up to 70 \degree C. KPS (25 mg; 0.093 mmol), previously dissolved in water (2,5 mL), was added after the emulsion was stirred for 15 min at 70 \degree C. After 4 h of stirring at 70 \degree C the reaction was stopped by contact with air. The resulting particles were directly filtered. The purification of the particles was carried out via centrifugation followed by decantation, addition of Milli-Q water and particle redispersion. The purification via centrifugation was performed three times. To remove SDS residues, the redispersed particles were dialysed in water for 24 - 72 h (dialysis tube membrane: 12 - 14 kDa). The particles were analysed with dynamic light scattering (ZetaSizer Nano DLS).

The PNIPAM-co-AA / PNIPAM-co-MAA shells were grown by first dissolving NIPAM(1.15g), BIS(11.5 mg) and MAA or AA(88.5 \textmu l) in 50 ml milli-Q water. Then 300 \textmu l of the core particle suspension was added. The mixture was stirred at 600 rpm for 20 min and purged with N\textsubscript{2}. The reaction mixture was heated up to 70 \degree C and KPS (25 mg; 0.093 mmol) was slowly added. After 2 h, the reaction reached completion and the microgel suspension was filtered and dialysed for 48h  (dialysis tube membrane: 12 - 14 kDa). The particles were characterized with DLS.

\subsection*{Fabrication of colloidal molecules} The responsive microswimmers were prepared using sequential capillarity-assisted particles assembly (sCAPA). The basic principles of sCAPA are described by Ni et al.\cite{Ni2016}. Here, we used an adapted version of sCAPA, which uses masters prepared with direct-writing via two-photon nanolithography, as opposed to Si masters made with conventional lithography methods. Briefly, negative masters were prepared with two-photon lithography (Nanoscribe Photonic Professional GT2) using an 63x oil-immersion objective and IP-Dip solution set.\cite{vanKesteren2022} The masters contained 6 smaller areas with 10,000 "traps" each, and one area was used to prepare the particles for one experiment. After printing and developing under standard conditions, these masters were post-cured for 2h under a UV-lamp and coated with a perfluorosilane using chemical vapor deposition. PDMS (Sylgard 184) was used to make templates from these masters. sCAPA was performed as described in the following paper \cite{Ni2015}. The PS particles were deposited from a 0.035\% Triton X-45 and 3.5 mM SDS solution in Milli-Q water. The microgels from a 0.04\% Triton X-45 in 1 mM HEPES pH 7.4 or 0.025\% Triton X-45 in 1 mM MES pH 4.5 for green and red-core microgels, respectively. All deposition were carried out at 25 $^\circ $ C with a deposition rate of 3 \textmu m s$^{-1}$.

\subsection*{Active motion experiments}

The colloidal clusters were activated using AC electric fields in a closed cell with the top and bottom electrodes consisting of transparent conductive slides separated by a 240 \textmu m-thick spacer (Grace Bio-Labs SecureSeal, USA, custom shape). The conductive slides were boro-silicate microscope slides coated with 3 nm Cr and 10 nm Au by thermal evaporation. The bottom slide on which the particles move was treated with thiolated polyethylene glycol (6 kDA, 1 mM in Milli-Q water, 30 min) to prevent sticking. The colloidal molecules were transferred from the sCAPA-templates to the top electrode, previously coated by a 3 \textmu m-thick spin-coated layer of glucose (40\% wt. in DI-water, 30 s 4000 rpm). Upon exposing the top slide to the suspending medium (0.1 mM HEPES buffer at pH 7 and 0.001\% wt. Triton in Milli-Q water) and closing the cell, the glucose layer dissolved and the particles sedimented to the bottom electrode. The conductive slides were connected to function generator (Agilent 33500B) with copper leads made from 50 \textmu m copper foil and the experiments were preformed on a Eclipse Nikon Ti2-e with a Lumencor Spectra II light box and a 20x objective. The particles were actuated by applying  an AC field with 800 Hz frequency and 7 V peak-to-peak amplitude. The field was applied for 15 min to equilibrate the sample before collecting data. Movies were recorded in a 1024*1024 px$^{2}$ window (340*340 \textmu m$^{2}$) at 10 frames per second. The local temperature was regulated through illumination with the 395 nm and 555 nm light sources of the light box (see SI for more details).

All the particles in the field of view were located and tracked using the TrackPy library in Python \cite{TrackPy}. Stuck particles and particles that get stuck during the experiment were omitted from the analysis. The instantaneous velocity of every individual particle was determined by measuring their displacements over 0.5 s windows (5 frames). The angular velocity was determined by a least-square fit the auto-correlation function of the direction of motion over 30 s windows (see SI for more details), corresponding to the time windows over which temperature was changed.

\bibliographystyle{unsrt} 
\bibliography{libPaper}

\end{document}

\maketitle

\makeatletter
\renewcommand{\thefigure}{S\@arabic\c@figure}
\makeatother
\tableofcontents

\newpage

\section{SI 1: Microgel characterization: temperature-dependent size, electrophoretic mobility}
The hydrodynamic size and eletrophoretic mobility of the microgels were measured on a Malvern Instruments Zetasizer Nano ZS. Microgels do not have a well-defined charged surface and therefore the electrophoretic mobility is reported instead of the $\zeta$-potential. The electrophoretic mobility experiments were performed in DTS1070 folded capillary cells with the microgels suspended in the same 0.1 mM HEPES buffer pH 7.4 as used in the AC-field driven active motion. The hydrodynamic radius is measured in a quartz cuvette in 0.1 mM HEPES buffer pH 7.4. The temperature was controlled by the Zetasizer.

\begin{figure*}[h]
\centering
\label{fig:FigS1}
\includegraphics[width=1\linewidth]{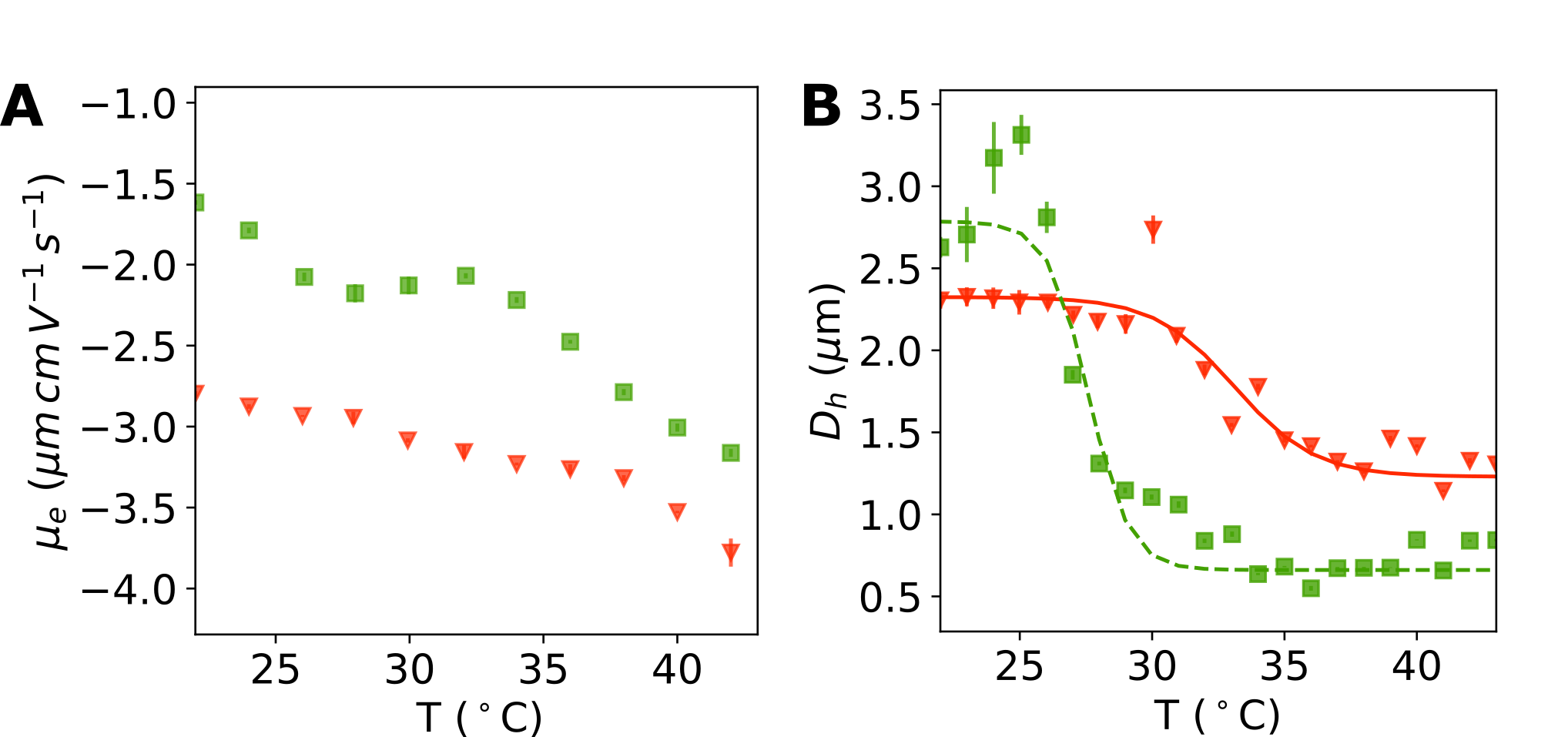}
\caption{A, Electrophorectic mobility of the G (squares) and R (triangles) microgels as a function of temperature. B, Hydrodynamic diameter of the G (squares) and R (triangles) microgels as a function of temperature. The error bars is the standard deviation of 5 measurements which sometimes fall within the symbol. }
\end{figure*}

\newpage

\section{SI 2: Estimation of the self-propulsion velocity of colloidal clusters}

As mentioned in the main text, in order to estimate the propulsion velocity of the clusters in their different states, we need to have access to their temperature-dependent size (as reported above in Section SI 1) and temperature-dependent dielectric properties. 

\subsection{Microgel characterization via dielectric spectroscopy}
The dielectric properties of the non-fluorescent PS particles and of the PNIPAM microgels were measured with a Novocontrol high-resolution dielectric analyzer (Alpha-A). The measurements were performed on a cell with a 6.45 mm gap distance between the electrodes enclosed by a Teflon cylinder, which was filled with the particle suspension at 0.1 wt\%. The permittivity ($\epsilon'$), and conductivity ($\sigma'$) were determined over a wide frequency range ($10^2$-$10^7$ Hz) and varying the temperature from 20\textdegree{}C to 55\textdegree{}C with a rate of  2\textdegree{}C/min (Supplementary Fig.~\ref{fig:DSfreq}). We further extracted $\epsilon'$ and $\sigma'$ corrected from the electrode polarization effect as function of temperature, at the frequency of 800 $Hz$, i.e. the one used in the experiments to actuate the  colloidal clusters (Supplementary Fig.~\ref{fig:DST}). We emphasize here that the dielectric spectroscopy measurements were carried out in fully de-ionized water, while the experiments are instead carried out in the presence of buffer and minute amounts of glucose remaining from the particle transfer. For this reason, the data reported below should only be considered appropriate to capture the trends in swimming velocity as a function of temperature and not as precise quantitative estimations. Nonetheless, we show that the predictions made from the measurements of dielectric properties in de-ionized water closely follow our experimental results, supporting our arguments. 
The dielectric spectroscopy measurements of a 0.1 \% w.t sample of the PS particles used to fabricate the colloidal assemblies, show a only a minor dependence of $\epsilon'$ with temperature and a linear increase in $\sigma$(Fig.S8), with no transition, as observed for the microgels.

\begin{figure*}[h]
\centering
\label{fig:DSfreq}
\includegraphics[width=0.95\linewidth]{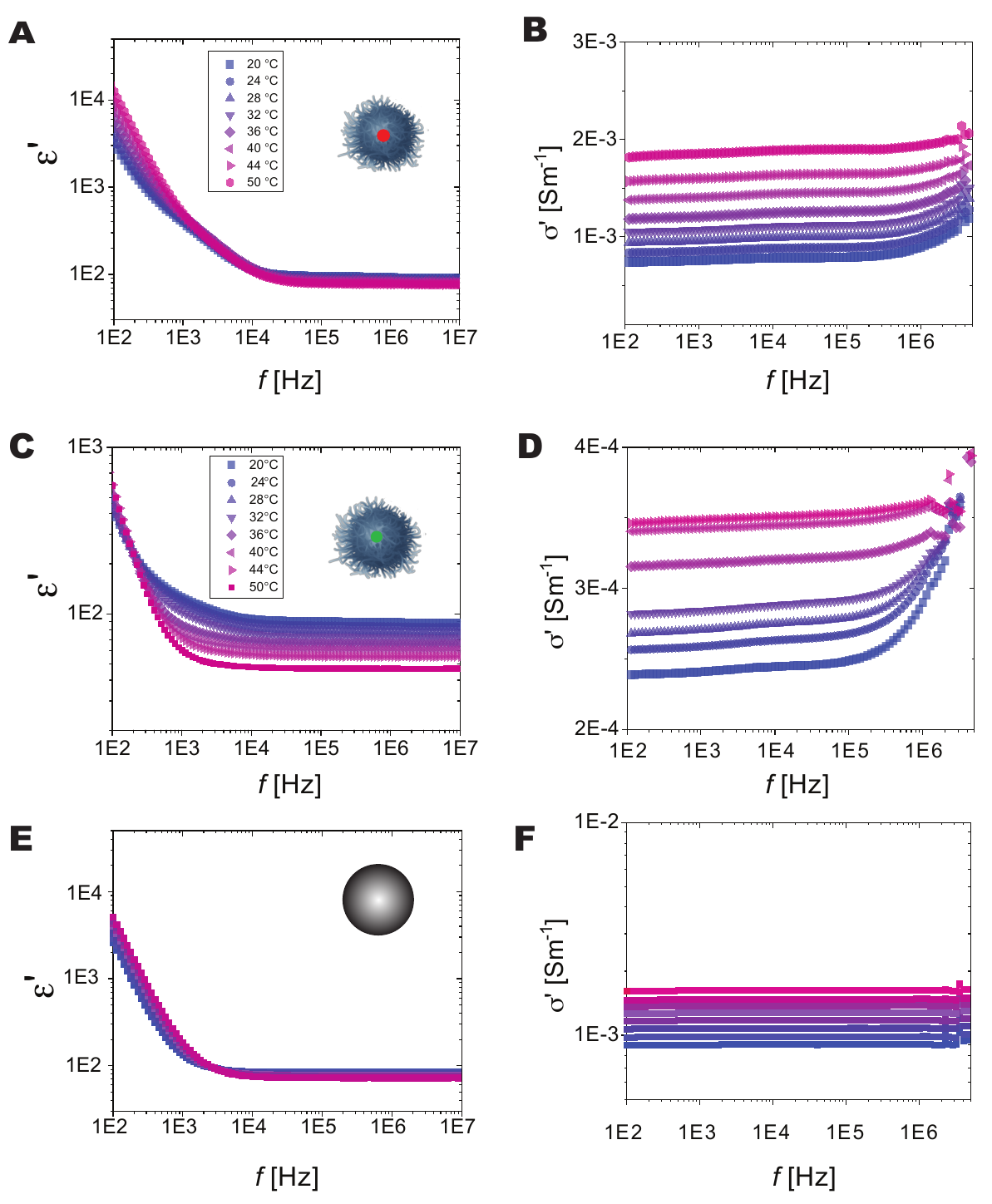}
\caption{\label{fig:DSfreq} $\epsilon'$ and $\sigma'$ as a function of frequency $f$ for an aqueous suspension at 0.1 wt \% of red microgels (A-B),  green microgels (C-D), and PS particles measured (E-F) for $T$ between $20$ and $50$\textdegree{}C.}
\end{figure*}

\begin{figure*}[h]
\centering
\label{fig:DST}
\includegraphics[width=\linewidth]{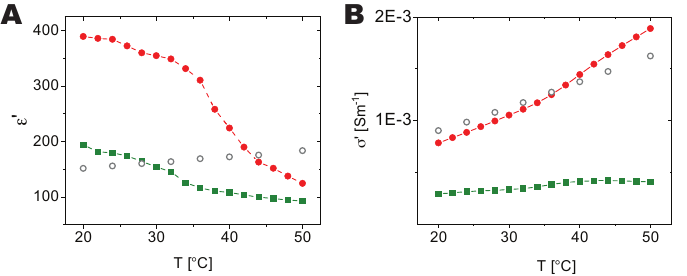}
\caption{\label{fig:DST} A, $\epsilon'$ and B, $\sigma'$ of the microgels (red and green solid symbols) and PS (gray open symbols) solution at 0.1 wt \% (corrected from the electrode polarization effect) from Fig.\ref{fig:DSfreq} as a function of temperature $T$  for R (red circles) and G (green squares) microgels at a fixed frequency of 800 Hz.}
\end{figure*}   

\subsection{Theoretical estimation of EHDFs and dumbbell velocity}
As described in previous work \cite{Alvarez2021}, the measured dielectric properties can be used to calculate the magnitude and direction of the EHDF velocities $U_{i}$ (Supplementary Fig.\ref{fig:UV}\textbf{A,B}), around a sphere under an AC electric field  described as \cite{Ma2015, Ma2015a}
\begin{equation}
\label{eq:EHD} U_i = \beta \frac{C}{\eta} \frac{K' + K''\bar{\omega}}{1+\bar{\omega}^2}\frac{3(r_i/R_i)}{2 \left[ 1+ (r_i/R_i)^2\right]^{5/2}},
\qquad
C = \epsilon \epsilon_0 H
\left( \frac{V_{\rm pp}}{2H} \right)^2;  
\end{equation}

where $\bar{\omega} = \omega H/ \kappa D$ with $\omega = 2\pi f$, $H$ twice the distance between the electrodes, $\kappa$ the Debye length, $D$ the diffusion coefficient of ions in solution and $\eta$ is the fluid viscosity. To evaluate the EHDF for each single lobe of a dumbbell, we consider $r_i$ as the distance from the center of the evaluated lobe to the center of the adjacent particle ($r = R_\textrm{\textmu{}gel}+ R_\textrm{PS} $). Therefore, as the radius of the microgel $R_\textrm{\textmu{}gel}$ varies with temperature, $r_i$ also changes. Finally, $K'$ and $K''$ are the real and imaginary part of the Clausius-Mosotti factor $K = K' + iK''$ \cite{CMFactor2017,Alvarez2021}. To calculate the Clausius-Mosotti factor, we extract the particle's $\epsilon_{p}$ and $\sigma_{p}'$ as described previously \cite{Alvarez2021} from the corrected $\epsilon'$ and $\sigma'$ as a function of temperature and obtain them as

\begin{equation}
    K' = \frac{\omega^2 (\epsilon_p - \epsilon_m) (\epsilon_p + 2\epsilon_m) + (\sigma_p - \sigma_m) (\sigma_p + 2\sigma_m)}{\omega^2 (\epsilon_p + 2 \epsilon_m)^2 + 2(\sigma_p + 2 \sigma_m)^2}
    \label{K1}
\end{equation}
\begin{equation}
    K'' = \frac{\omega (\epsilon_p - \epsilon_m) (\sigma_p + 2\sigma_m) - (\epsilon_p + 2\epsilon_m) (\sigma_p - \sigma_m)}{\omega^2 (\epsilon_p + 2 \epsilon_m)^2 + 2(\sigma_p + 2 \sigma_m)^2}
    \label{K2}
\end{equation}

where the subscript $_{m}$ refers to the medium and $_{p}$ refers to the particle.

Combining the $T$-dependent velocities of the EHDFs for each particle $U_i$, the dumbbell velocity $v_{\mu gel, PS}$ can, in first approximation, be obtained as a linear combination of the two values of $U_{\rm i}$ (Supplementary Fig.\ref{fig:UV}\textbf{C}) as 

\begin{equation}
\label{eq:vij} v_{\mu gel, PS}=U_{\mu gel}r_{PS}+U_{PS}r_{\mu gel}/(r_{\mu gel}+r_{PS})
\end{equation}

\begin{figure*}[h]
\centering
\label{fig:UV}
\includegraphics[width=\linewidth]{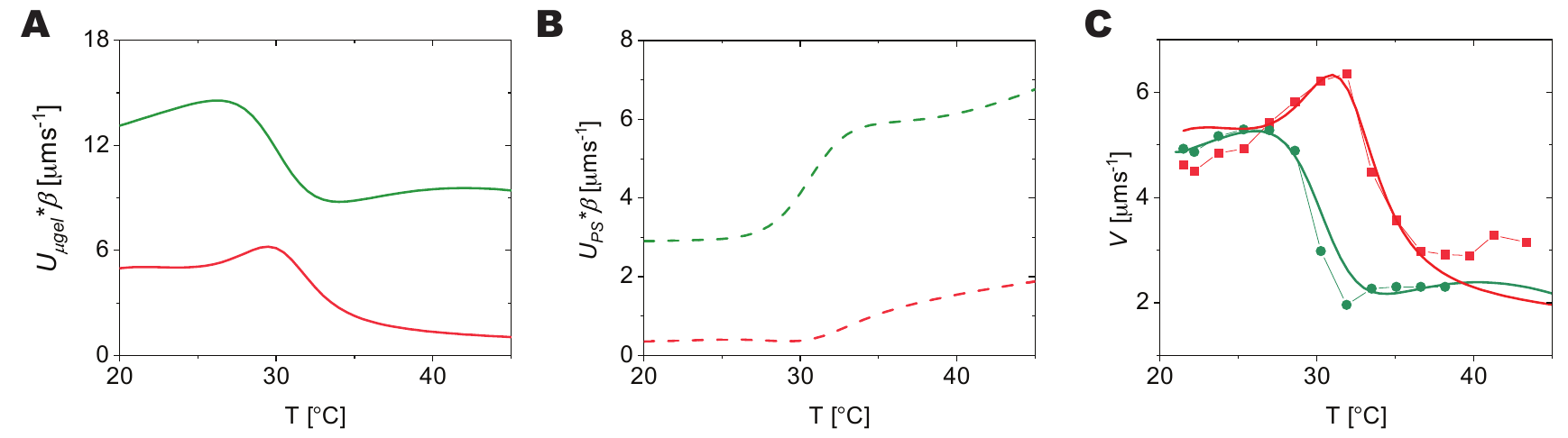}
\caption{\label{fig:UV} EHDF velocities $U_i$  calculated from Eq. (\ref{eq:EHD}) for A, Red (red solid line) and green (green solid line) microgels ($U_{\mu gel}$), and B, for the PS particle $U_{PS}$ close to each microgel (green and red dashed lines). The final $U_i$ value is obtained including the $\beta$ prefactor as one single free parameter $\beta_{PS}=0.01$ $\beta_{R}=0.075$ for the R-PS dumbbell and $\beta_{PS}=0.07$ $\beta_{G}=0.23$  for the G-PS dumbbell. C,  Experimental values (symbols) and theoretical estimation (solid line) of dumbbell velocity as a function of temperature calculated from Eq. (\ref{eq:vij}). }
\end{figure*}

\clearpage
\section{SI 3: Temperature control with plasmonic heating: Experiments and FEM simulations}

\subsection{Illumination and absorption characterization}
All experiments where performed on a NIKON Eclipse TI2-e microscope with a Lumencor Spectra II light source. To determine the intensity of the light leaving the microscope objective under different conditions, a Thorlabs S170C power sensor photo-diode in combination with a PM100D console was used. The sensor only measures the total light intensity and therefore was passed through a 0.0125 mm\textsuperscript{2} pinhole to get the power density. Assuming a homogeneous illumination, the light intensity without the pinhole was used to calculate the area of the light spot. The results are summarized in Table \ref{tab:TabS1.1}. The transmittance of the Au layer was determined using the power density measured with an Au-coated slide between the objective and the sensor. The results are shown in Table \ref{tab:TabS1.2}. These illumination parameters are used in the Finite Element Method (FEM) model. Only the 20x objective with 395 nm and 555 nm illumination lines are used in experiments and only these parameters are modelled further. The light intensity for each wavelength can be controlled between 1 and 100 \% using the Nikon Imaging Software and these values are calibrated to give the corresponding power densities. \\

The heating due to absorption of the Au layer was measured using a small thermo-couple connected to an OKO-lab temperature controller, which logs the temperature in 1 second intervals. The thermo-couple was placed in the cell between the conductive Au-coated glass slides through the filling hole and the cell was filled with milli-Q water. The thermal gradient was measured by centering the objective first on the thermo-couple and subsequently moved in controlled steps from this position with the motorized stage. The ambient temperature was logged with a separate thermo-couple and the heating is obtained as the difference between the temperature in the cell and the ambient temperature.

\begin{table}[h]
\centering
\begin{tabular}{|p{2cm}|p{2cm}|p{2cm}|p{2cm}|p{2cm}|  }
\hline
Objective   & Spot size (mm\textsuperscript{2}) & PD 640 nm  (mW mm\textsuperscript{-2}) & PD 555 nm (mW mm\textsuperscript{-2}) & PD 395 nm (mW mm\textsuperscript{-2}) \\
\hline
4x NA 0.13  & 13.73                             & 3.2                                               & 3.6                                              & 2.9                                              \\
\hline
20x NA 0.45 & 0.55                              & 48.0                                              & 58.9                                             & 45.0                                             \\
\hline
40x NA 0.6  & 0.16                              & 89.1                                              & 110.0                                            & 86.2                                             \\
\hline
60x NA 0.7  & 0.07                              & 109.2                                             & 139.9                                            & 107.9         \\
\hline
\end{tabular}
\caption{\label{tab:TabS1.1} Power densities and spot size for different objectives}
\end{table}

\begin{table}[h]
\centering
\begin{tabular}{|p{2cm}|p{2cm}|p{2cm}|p{2cm}|}
\hline
Wavelength (nm) & Transmission (\%) & Absorption (\%) & Reflection (\%) \\ \hline
395             & 27                & 55             & 18              \\ \hline
555             & 38                & 41             & 21              \\ \hline
640             & 37                & 35             & 28              \\ \hline
\end{tabular}
\caption{\label{tab:TabS1.2} Transmission, Absorption and Reflection of the 10 nm Au-layer}
\end{table}

\subsection{FEM modeling}
\begin{figure}[h]
\label{Fig S1.1}
\includegraphics[width=\linewidth]{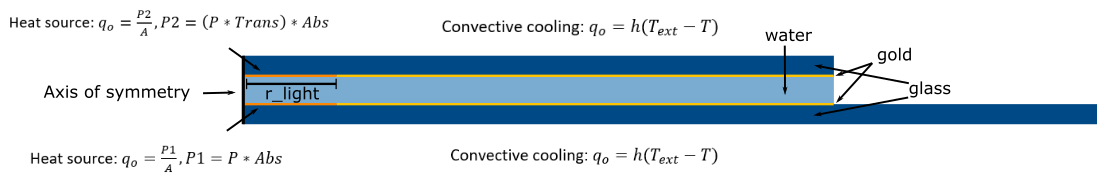}
\caption{\label{fig:FigS1.1} Schematic description of the COMSOL model. The dimension are not in scale. }
\end{figure}

\begin{figure*}[h]
\label{Fig S1.2}
\includegraphics[width=\linewidth]{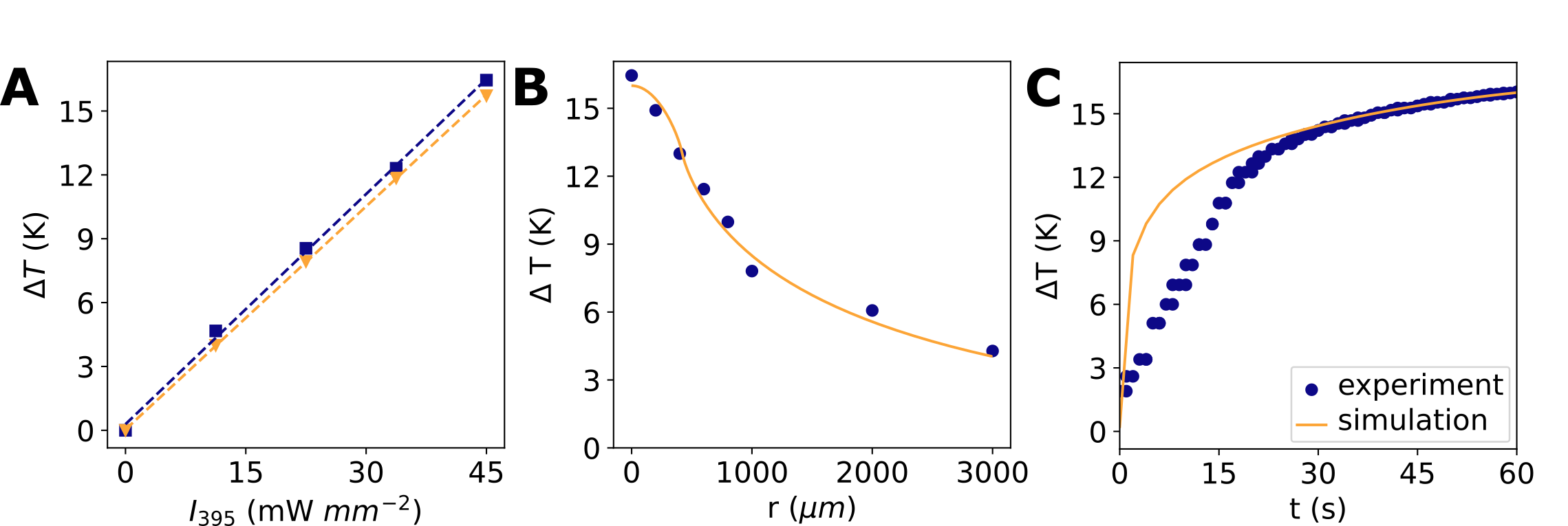}
\caption{\label{fig:FigS1.2} Comparison between experiments and simulations. A, Heating as a function of light intensity, measured with the thermo-couple (squares) and modelled with COMSOL (triangles), in the center of the illuminated spot after 60 s for the 395 nm line. B, Thermal profile measured from the center of the illuminated spot (r = 0) in experiments (circles) and simulations (solid line) after 60 s at a power density of 45 mW mm\textsuperscript{-2}. C, Time-dependent heating measured experimentally (circles) and modelled (solid line) at the center of the illuminated spot for a power density of 45 mW mm\textsuperscript{-2}.}
\end{figure*}

The FEM model is constructed in COMSOL Multiphysics \cite{COMSOL} to be an accurate description of our setup, but simplifications are made to ease computation. A schematic overview is shown in Figure \ref{fig:FigS1.1}. The model is a time-dependent FEM 2D simulation with axial symmetry and uses the in-build "Heat transfer in solids and Fluids (ht)", "Laminar flow (spf)" and "Nonisothermal Flow (nitf1)" physics simulations from COMSOL. Therefore, both the diffusive and convective heat transfers are modelled. The plasmonic heating in the thin gold films is not directly modeled, but we accounted for it in a simplified way by including heat sources over an area corresponding to the light spot on both the top and bottom slides. The magnitude of the heat source on the bottom is determined by the power density of the light and the expected absorption of the Au layer (the values for 395 nm are used unless stated otherwise). The heat source on the top is determined from the power density that is transmitted through the bottom Au layer taking both absorption and reflection into account. The top and bottom slides are expected to be cooled via convection. The edge of the bottom slide is in direct contact with the metal of the microscope stage which acts as a heat sink and is modelled as a constant temperature. The model is compared to values measured with a temperature probe and the sensitivity to mesh size and time step is also characterized. \\ 
The FEM model has an excellent agreement with experiments in both the power scaling (Figure \ref{fig:FigS1.2}A) and the thermal gradients (Figure \ref{fig:FigS1.2}B). There is a discrepancy in the initial time-dependent heating (Figure \ref{fig:FigS1.2}B), however this is likely a limitation of the experimental method due to the non-negligile thermal mass of the thermal probe.   

\begin{figure*}[h]
\centering
\label{Fig S1.3}
\includegraphics[width=\linewidth]{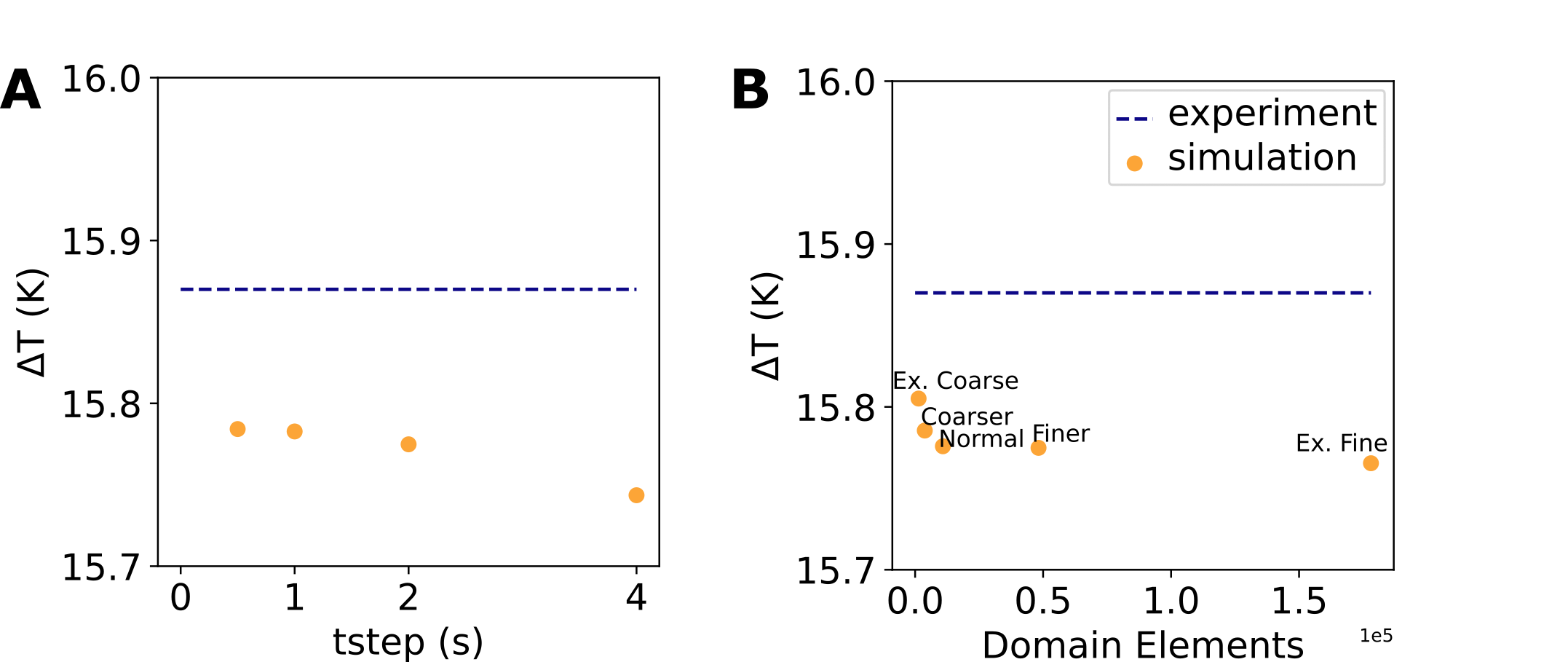}
\caption{\label{fig:FigS1.3} Sensitivity analysis of the COMSOL model for modelling parameters. A, Analysis of the sensitivity as a function of the time step for the heating in the center at 45 mW mm\textsuperscript{-2} and t = 60 s. B, Analysis of the sensitivity a a function of the mesh size for the final heating in the center at 45 mW mm\textsuperscript{-2} and t = 60 s. The text above the data points are the mesh settings used in COMSOL. The dashed lines mark the experimentally measured heating. }
\end{figure*}

\clearpage

\newpage

\section{SI 4: 3D-printed trap design for sequential capillarity-assisted particle assembly of microgels}

\begin{figure*}[h]
\centering
\label{Fig S3.1}
\includegraphics[width=1\linewidth]{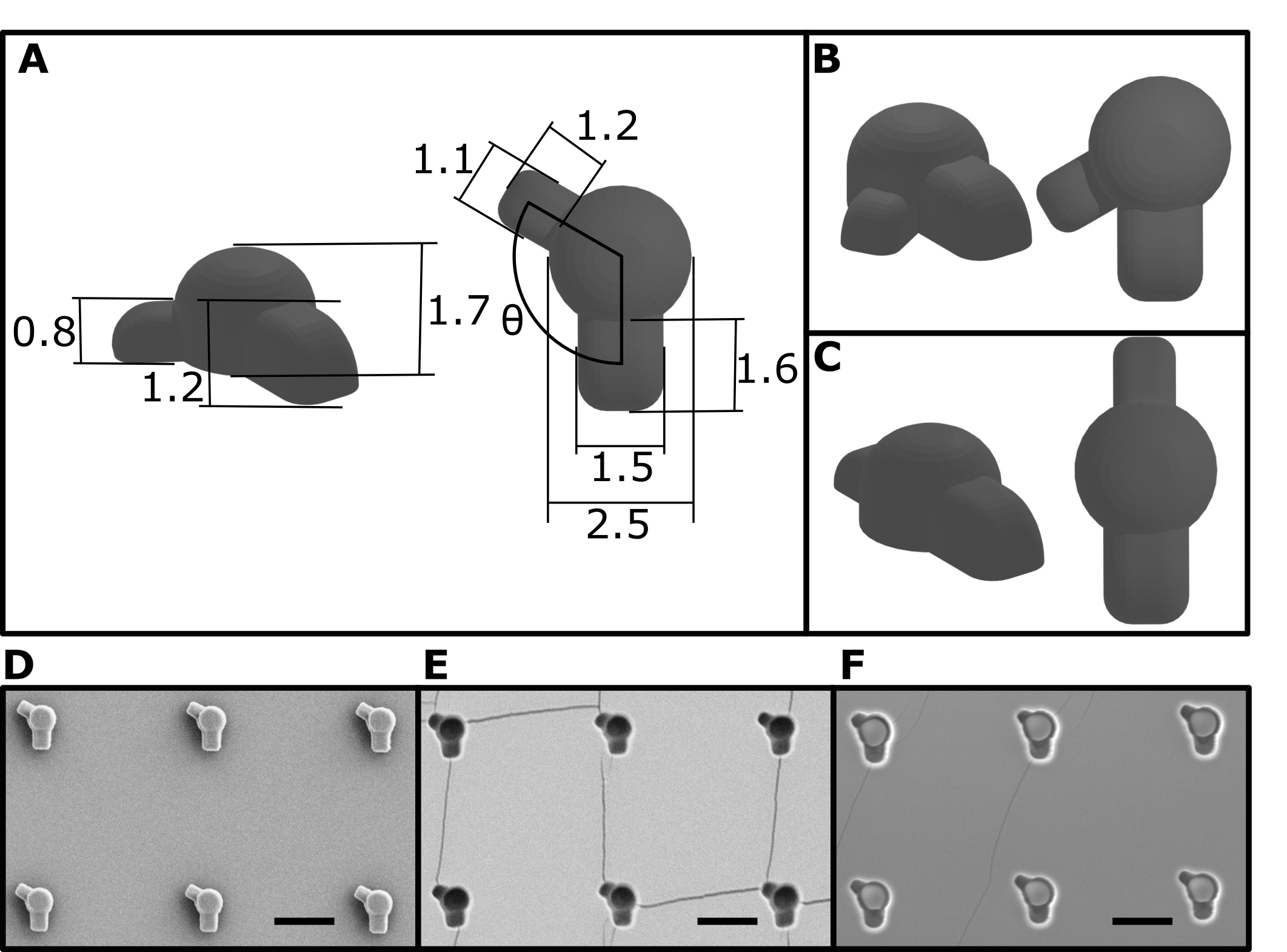}
\caption{ A, Side and top view of the 3D-model for the sCAPA-masters for the 120-cluster. The dimensions for the different parts are given in \textmu m. B, Side and top view of the 3D-model for the sCAPA-masters for the 60-cluster. C, Side and top view of the 3D-model for the sCAPA-masters for the 180-cluster. D, SEM-micrograph of the 3D-printed sCAPA-masters for the 120-cluster. E, SEM-micrograph of the 3D sCAPA-traps for the 120-cluster. F, SEM-micrograph of the 3D sCAPA-traps for the 120-cluster filled with the PS-particles. Scale bars are 5 \textmu m }
\end{figure*}

\clearpage
\section{SI 5: Yield of the sCAPA of R-G-PS clusters }

\begin{figure*}[h]
\centering
\label{fig:FigS3.2}
\includegraphics[width=\linewidth]{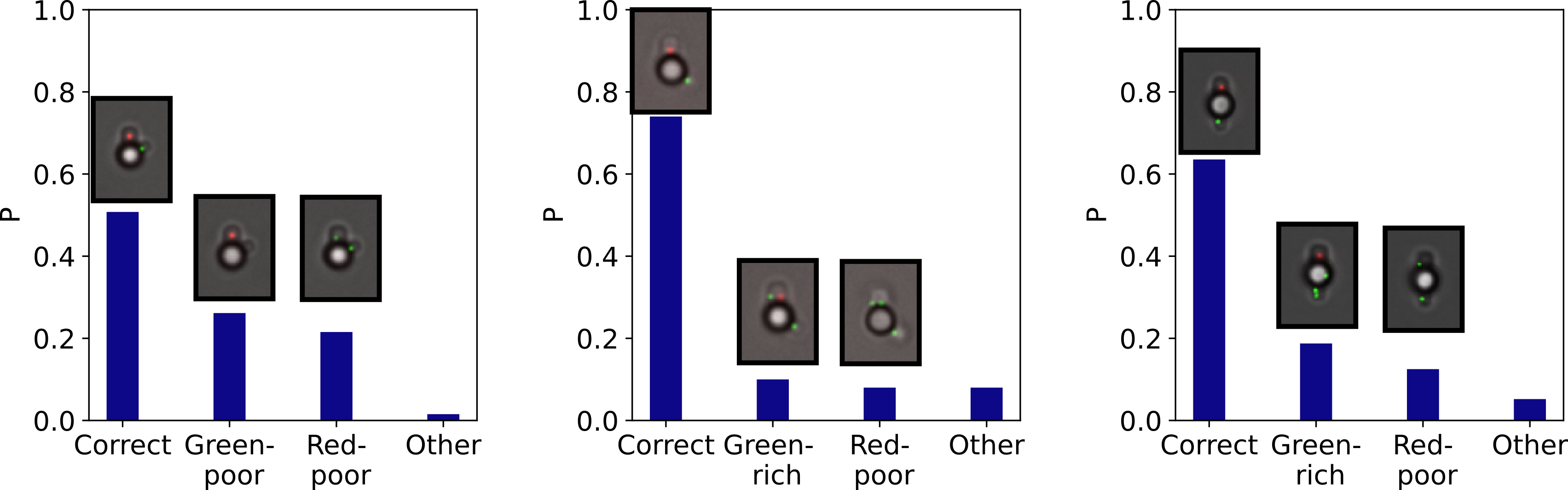}
\caption{ Yields of the sCAPA for A, 60 cluster B, 120-clusters, and C, 180-clusters. The inset images are representative of the most common errors for the different depositions. These errors are a surplus of green-microgels in the wrong places (green-rich) or a lack of red or green microgels (red-poor or green-poor)}
\end{figure*}

\clearpage

\section{SI 6: Determination of the angular velocity from trajectories}

\begin{figure*}[h]
\label{fig:FigS5.1}
\centering
\includegraphics[width=\linewidth]{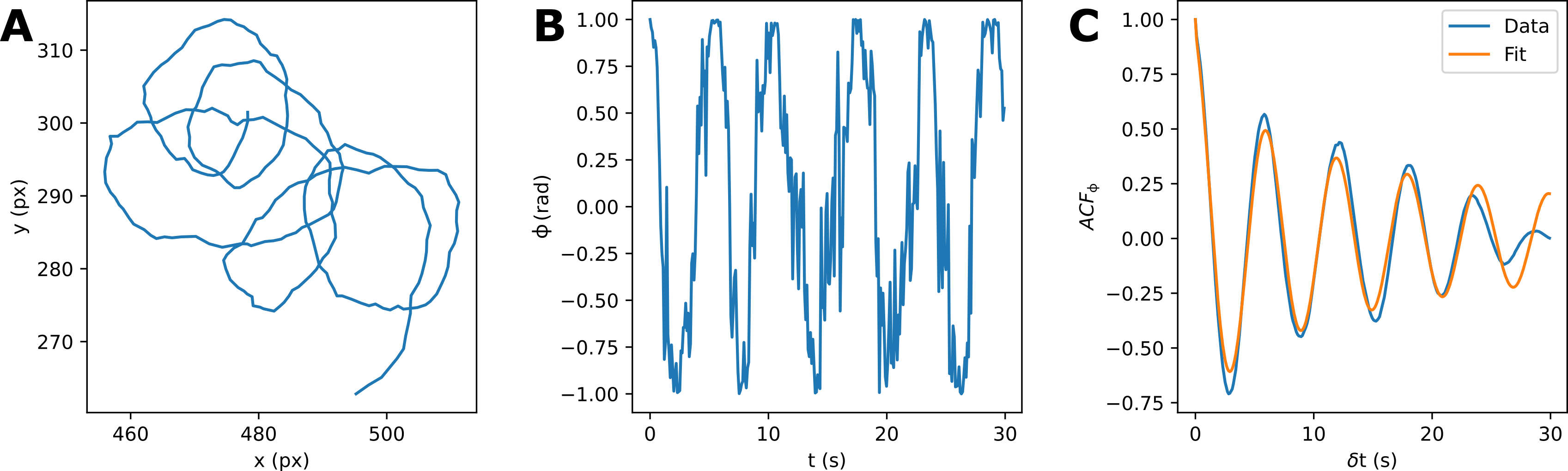}
\caption{ A, An example particle trajectory. B, The direction of motion over time for the trajectory shown in A. C, The auto-correlation function and t least-square fit.}
\end{figure*}

The angular velocities are determined using a fit of the auto-correlation function of the direction of motion ($\phi$).  (Figure \ref{fig:FigS5.1}A). At each time step, the direction of motion is determined from ratio of the $x$ and $y$ displacements as $\phi = arctan(\frac{\Delta y}{\Delta x})$, with $\Delta x = x_{t+\Delta t}- x_t$ and $ \ \Delta y = y_{t+\Delta t}- y_t$ (Figure \ref{fig:FigS5.1}B). We then calculate the time auto-correlation function of $\phi$, normalize it to its initial value and perform a least-square fit with the following function: $ACF_{\phi} (\delta t) = cos(\omega* \delta t)*e^{-D_r\delta t}$ (Figure \ref{fig:FigS5.1}C) \cite{Mano2017}.\\

To validate the determination of the angular velocity and to errors, we simulate active Brownian particles with chiral motion. The model follows as: 

\begin{align*} 
\centering
\dot x(t) = v cos(\phi(t))+ \sqrt{2D_t}\xi,\\
\dot y(t) = v sin(\phi(t))+ \sqrt{2D_t}\xi,\\
\dot \theta(t) = \omega + \sqrt{2D_r}\xi,\\
\end{align*}
\\

With $v$ as the instantaneous velocity, $D_t$ the translational diffusion, $D_r$ the rotational diffusion, $\omega$ the angular velocity and $\xi$ a Gaussian white noise term. We simulate 100 particles for 30 s with a time step of 0.1 s. These values are comparable with the experimental conditions. We then extract the time auto-correlation function from the simulated trajectories and compare the results of the fit with the input angular velocity. We test the sensitivity for the rotational diffusion, angular velocity and instantaneous velocity. The translational diffusion was fixed at 0.15 \textmu m\textsuperscript{2} s\textsuperscript{-1} which is the diffusion of a 3 \textmu m colloid in water.

\begin{figure*}[h]
\label{Fig S5.2}
\includegraphics[width=\linewidth]{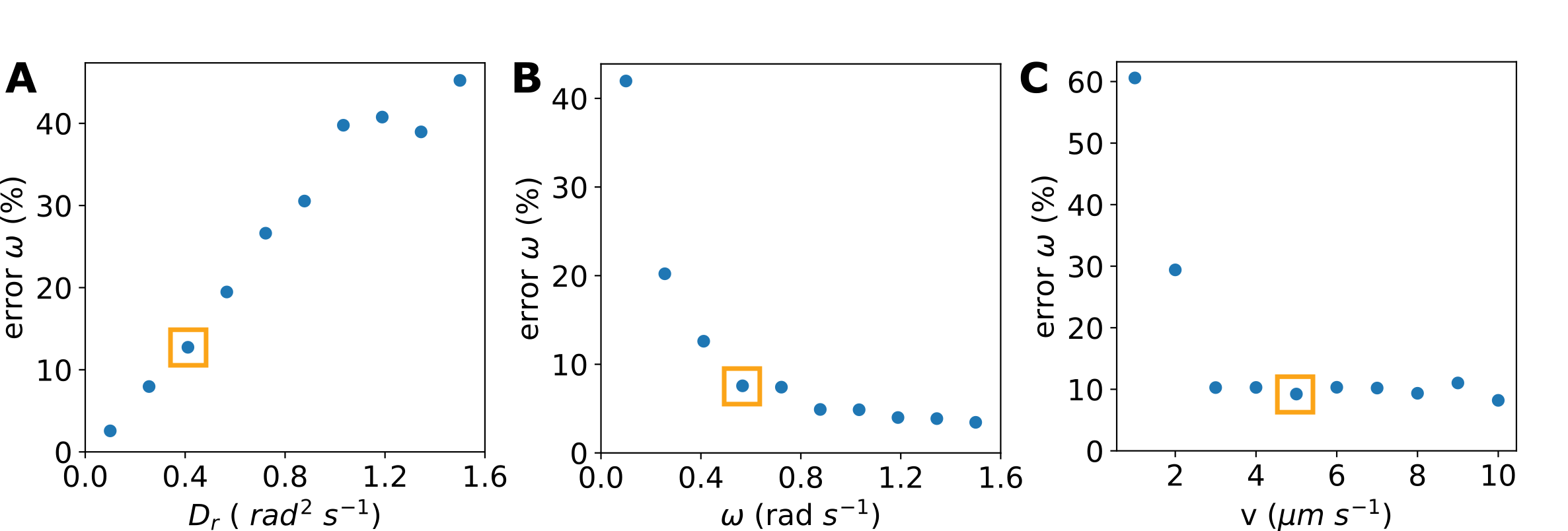}
\caption{\label{fig:FigS5.2} Sensitivity analysis of the angular velocity determination algorithm as a function of $D_{R}$, 
$\omega$ and $v$. In each of the parameter sweeps one parameter is varied and the other two are fixed to the value corresponding to the yellow square.  A) Mean absolute error percentage of $\omega$ as a function of rotational diffusion. B) Mean absolute error percentage of $\omega$ as a function of angular velocity. C) Mean absolute error percentage of $\omega$ as a function of instantaneous velocity.}
\end{figure*}

In Figure \ref{fig:FigS5.2}, the results of this validation are shown. We observe that the error increases roughly linearly with the rotational diffusion (Fig. \ref{fig:FigS5.2}A).  Moreover, errors are larger for lower angular velocities (Fig. \ref{fig:FigS5.2}B) due to the increasingly more limited statistics over the time frame of the experiment or simulation. Interestingly, the algorithm is relatively insensitive to the instantaneous velocity except for the low-velocity range (Fig. \ref{fig:FigS5.2}C). This stems form the fact that, for low propulsion velocities, translational diffusivity plays an increasingly more dominant role in determining displacements and the expression used to fit the auto-correlation function no longer holds. Comparing these results with the values reported in the main text, we note that errors in the estimation of $\omega$ are typically below 20\%, and most often below 10\%.

\clearpage

\section{SI 7: Estimation of the swimming velocity for R-G-PS trimers with different opening angles}

We use a simple triangular vector model to estimate the final swimming velocity $\textbf{v}$ for a trimer with internal angle $\theta$. We decompose the propulsion of the trimer into the the one generated by each PS-microgel dumbbell and consider their linear superposition (Fig.\ref{fig:triangle}\textbf{A}). In particular, we obtain the two in-plane velocity components for each dumbbell as

\begin{equation}
\label{eq:Vtriangle} \textbf{v}_{x,i} = \textit{v}(T)  \frac{cos\theta}{2} \textbf{\hat{x}};
\end{equation}

\begin{equation}
\label{eq:Vtriangle} \textbf{v}_{y,i} = \textit{v}(T)
\frac{sin\theta}{2} \textbf{\hat{y}};
\end{equation}

where \textit{v(T)} is the velocity magnitude as a function of temperature for each dumbbell, obtained from Fig.\ref{fig:UV}\textbf{C} for the experimental and theoretical data, $\theta$ the opening angle between the microgels attached to a the PS particle, and $\textbf{\hat{x}}$ and $\textbf{\hat{y}}$ are the unit vectors in x and y, respectively. Finally, we sum the \textbf{v}$_{x,i}$ and \textbf{v}$_{y,i}$ components to obtain the trimer velocity $\textbf{v}$ as:

\begin{equation}
\label{eq:Vfinaltrimer} \textbf{v} = [(\textbf{v}_{x,R} + \textbf{v}_{x,G}), (\textbf{v}_{y,r} + \textbf{v}_{y,g})].
\end{equation}

We finally calculate the magnitude of $\textbf{v}$ as a function of the temperature $T$ for each case as $\lVert  \textbf{v}\rVert= \sqrt{(\textbf{v}_{x})^{2} + (\textbf{v}_{y})^{2}}$ (Fig.\ref{fig:triangle}\textbf{B}).

\begin{figure*}[h]
\centering
\label{fig:triangle}
\includegraphics[width=0.8\linewidth]{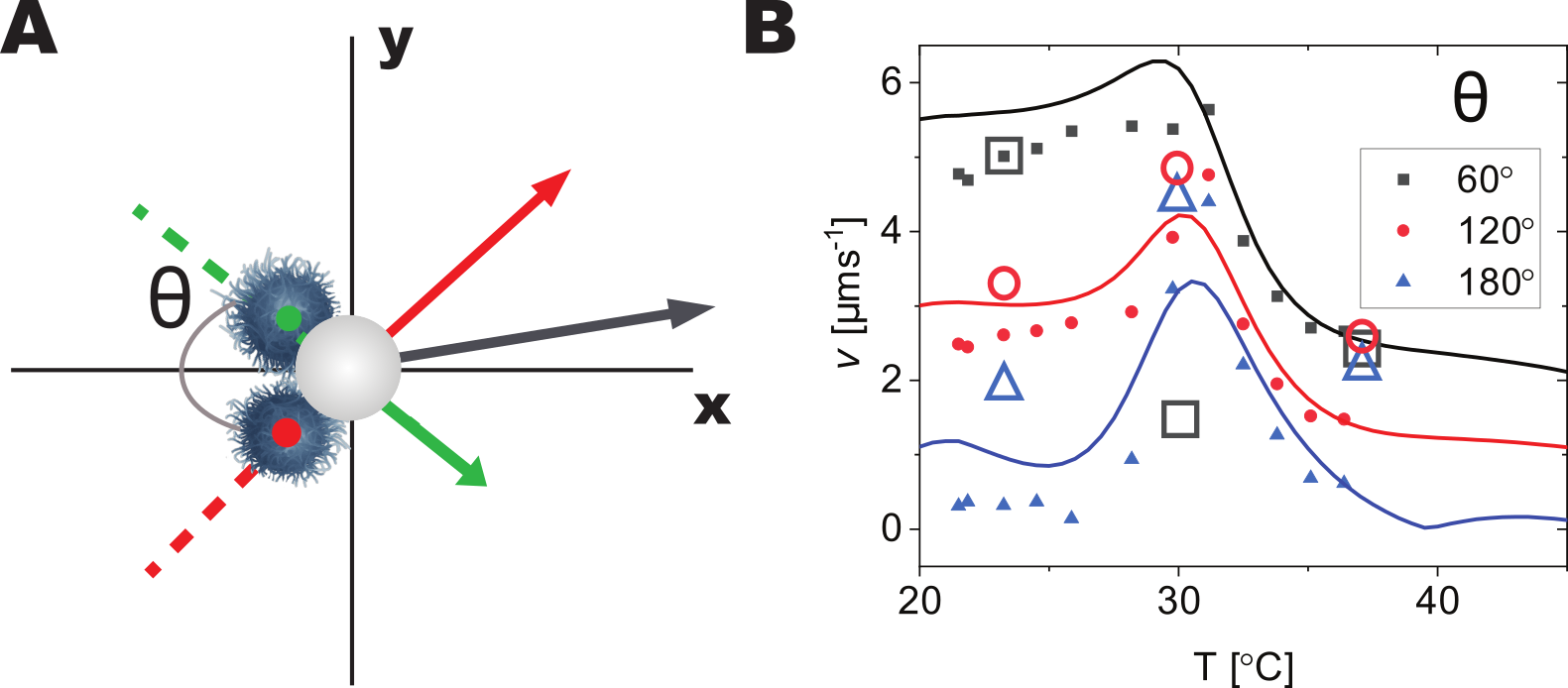}
\caption{\label{fig:triangle} A, Scheme of a R-G-PS trimer, with an opening angle $\theta$ between the microgels. The green and red arrows represent the velocity vectors for each PS_$\mu gel$ pair, obtained from the data fitted in Fig.\ref{fig:UV}. The black arrow indicates the sum velocity vector. B, Estimated trimer velocities $v$ as a function of temperature for each experimental configuration with  $\theta$ 60 $\degree$, 120 $\degree$, and 180 $\degree$ (black, red and blue respectively). The symbols indicate the result using the experimental dumbbell velocity, and the solid lines are fits to the data, using the vectorial sum of the expressions obtained from Eq. \ref{eq:vij} for each dumbbell. The large open symbol represent the experimental velocities of the different clusters.}
\end{figure*}

\clearpage

\section{SI 8: Calculation of microgel overlap on R-G-PS trimers}

\begin{figure*}[h]
\label{Fig S6.1}
\includegraphics[width=\linewidth]{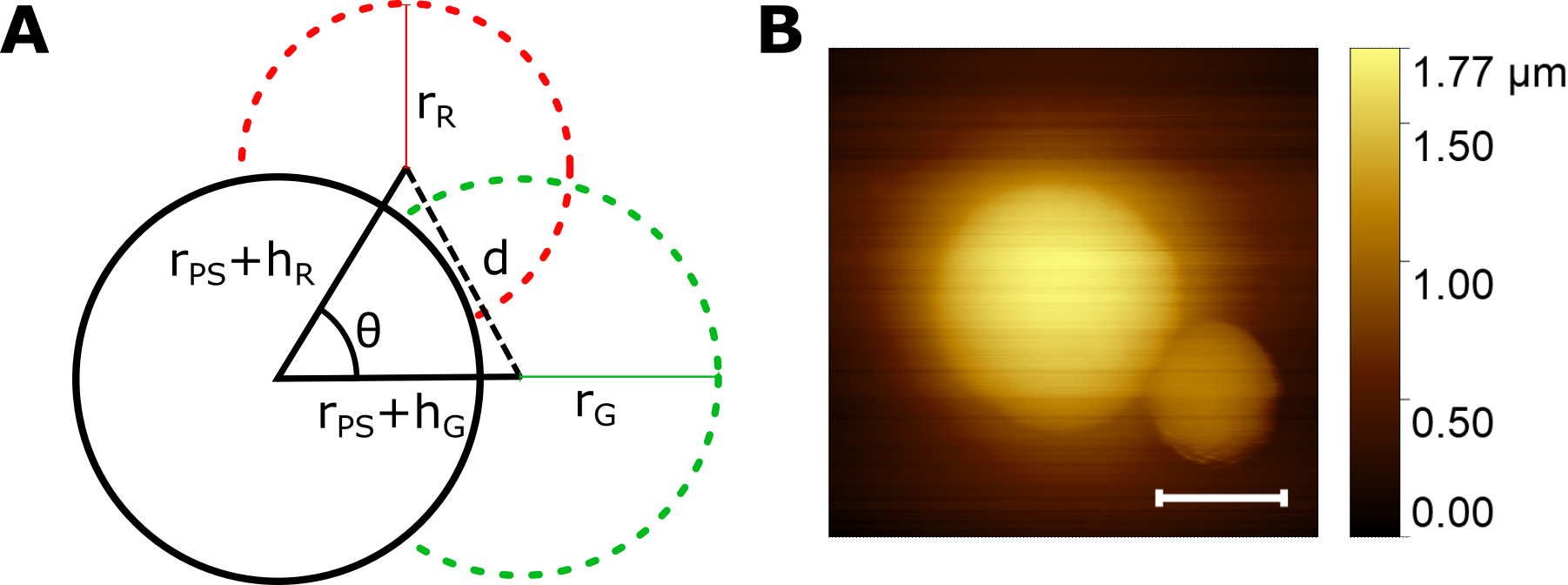}
\caption{\label{fig:FigS6.1} Calculation of the theoretical overlap between the microgels adsorbed on a PS particle. A, Schematic of the geometry of t eh problem used for the calculation. B, AFM micrograph of a R-PS dumbbell in water. Scale bar is 1 \textmu m}

\end{figure*}

Here, we report simple geometrical calculations to estimate the overlap of the two microgels attached to the PS particle at different opening angles $theta$. Based on the scheme in  Figure \ref{fig:FigS6.1}A, we obtain the following relationships:
\begin{gather}
\label{eq:overlap} 
h_{R} = h_{G} = h \\
d = 2r_{PS}*sin(\frac{1}{2}\theta)+2h*sin(\frac{1}{2}\theta)
\end{gather}

Here, $r_{PS}$ is the radius of the hard PS colloid (1 \textmu m), $r_{R}$ is the radius of the red-core microgel, $r_{G}$ is the radius of the green-core microgel and  $h_{R}$ and $h_{G}$ are the distances at which the microgels' centers sit from the PS surface. The radii are easily found with the DLS data (Figure \ref{Fig S1}B) and are 1.15 \textmu m and 1.3 \textmu m for the G and R microgels, respectively. However, the height from the surface is more difficult to determine. Here we use atomic force microscopy (AFM) in water to probe the configuration of the microgel on the particle surface (Figure \ref{fig:FigS6.1}B). The AFM does not probe the full microgel as the sharp tip cannot easily detect the loosely cross-linked polymer chains in the corona at the periphery of the particle and only the more-densely cross-linked core is clearly visible. However, AFM images can still be used to measure the distance $h$. We assume that $h_{R} = h_{G} = h$ measure a value of approximately 400 nm. \\

From these simple geometrical considerations, we find that our microgels start to overlap at an opening angle $\theta \lesssim 120^{\circ}$. This confirms that there is significant overlap for the 60-clusters, while no overlap is found for the 120- and 180-clusters. These results indicate some simple design guidelines to achieve multiple states for particles and microgels of different sizes.

\newpage

\bibliographystyle{unsrt} 
\bibliography{libPaper}